\documentclass[journal,twoside]{IEEEtran}
\usepackage{amsmath}

\usepackage{amssymb}

\newtheorem{definition}{Definition}
\newtheorem{lemma}{Lemma}
\newtheorem{theorem}{Theorem}
\newtheorem{proposition}{Proposition}
\newtheorem{remark}{Remark}

\newcommand{\N}{{\mathbf N}}
\newcommand{\R}{{\mathbf R}}
\newcommand{\C}{{\mathbf C}}
\newcommand{\diag}[1]{\textnormal{diag}\left(#1\right)}
\newcommand{\Mat}[2]{\textnormal{M}_{#1}\left(#2\right)}
\newcommand{\CvM}[2]{\textnormal{CM}_{#1}}
\newcommand{\calS}{\mathcal{S}}
\newcommand{\calU}{\mathcal{U}}
\newcommand{\Tr}[1]{\textrm{Tr}\left(#1\right)}
\newcommand{\tr}[1]{\textrm{tr}\left(#1\right)}
\newcommand{\TR}{\textrm{tr}}
\newcommand{\E}[1]{{\mathbf E}\left(#1\right)}
\newcommand{\calA}{\mathcal{A}}
\newcommand{\F}[1]{\varphi\left(#1\right)}
\newcommand{\FI}{\varphi}
\newcommand{\calG}{\mathcal{G}}
\newcommand{\class}[1]{\mathfrak{#1}}

\begin{document}
\title{On the Capacity of Block Multiantenna Channels}
\author{Mario~Diaz
        and Victor P\'{e}rez-Abreu%
\thanks{Mario Diaz is with the Department of Mathematics and Statistics at Queen's University, Canada. See \textit{www.mariodiaztorres.com} for current contact information.}%
\thanks{Victor P\'{e}rez-Abreu is with the Probability and Statistics Department at the Center for Research in Mathematics (CIMAT), Mexico. E-mail: \textit{pabreu@cimat.mx}.}%
\thanks{Copyright (c) 2017 IEEE. Personal use of this material is permitted.  However, permission to use this material for any other purposes must be obtained from the IEEE by sending a request to pubs-permissions@ieee.org.}}

\markboth{}%
{DIAZ AND P\'{E}REZ-ABREU: ON THE CAPACITY OF BLOCK MULTIANTENNA CHANNELS}


\maketitle

\begin{abstract}
In this paper we consider point-to-point multiantenna channels with certain block distributional symmetries which do not require the entries of the channel matrix to be either Gaussian, or independent, or identically distributed. A main contribution is a capacity theorem for these channels, which might be regarded as a generalization of Telatar's theorem (1999), which reduces the numerical optimization domain in the capacity computation. With this information theoretic result and some free probability arguments, we prove an asymptotic capacity theorem that, in addition to reducing the optimization domain, does not depend on the dimension of the channel matrix. This theorem allows us to apply free probability techniques to numerically compute the asymptotic capacity of the channels under consideration. These theorems provide a very efficient method for numerically approximating both the capacity and a capacity achieving input covariance matrix of certain channels.
\end{abstract}


\section{Introduction}
\label{Section:Introduction}
\IEEEPARstart{R}{ight} after the introduction of multiantenna systems in the late 1990s \cite{FoGa98,Fo96}, Random Matrix Theory (RMT) became an important tool for the analysis of these systems, as evidenced in \cite{TuVe04}. A connection between multiantenna communications and RMT arose from the relation between the capacity of a point-to-point multiantenna channel and the eigenvalue distribution of the matrix of propagation coefficients of the system. Among other things, in his pioneering paper \cite{Te99}, Telatar found the asymptotic normalized ergodic capacity in the number of antennas of a point-to-point multiantenna channel with independent identically distributed (i.i.d.) Rayleigh faded propagation coefficients, using one of the most celebrated results in RMT, the Marchenko--Pastur theorem \cite{MaPa67}. With this asymptotic result, and the fact that the asymptotic normalized capacity provides an approximation for the corresponding non-asymptotic quantity, large capacity gains for multiantenna systems compared to their single antenna counterparts were predicted. In general, RMT asymptotic approximations become more accurate as the number of antennas in the system grows. Thus, in applications where the number of antennas is large, e.g., a possible link between two massive antenna arrays like the ones in \cite{EdLaMaTu14}, a very close correspondence between the real performance of the system and the asymptotic prediction is expected.  Since the assumptions of independence, equal distributions, and Rayleigh fading might be restrictive, models that relax some of these hypotheses have been considered.

In most practical cases, the Rayleigh fading assumption can be dismissed, due to the universal behavior exhibited by some spectral statistics of certain ensembles of random matrices \cite{BaSi10,TaVu12,TaVu12_2,Di14,GoKoTi15}. A common alternative to assuming equal distributions is to require the entries to be independent and their variances to follow a certain pattern. Examples of channels in this category are the IND channels in Tulino et al. \cite{LoTuVe05}. In fact, their paper considers the more general Kronecker correlation model, which, after some transformations, can be studied by means of IND channels. For the computation of the asymptotic channel capacity, a key step is to compute the deterministic equivalent of the empirical Stieltjes transform of the channel matrix, a technique introduced by Girko \cite{Gi90,Gi01}. In general, deterministic equivalents are computed by solving a non-linear system of equations, which in the i.i.d. Rayleigh fading case reduces to a quadratic functional equation satisfied by the Stieltjes transform of the Marchenko--Pastur law, see \cite{HaLoNa07} and the references therein for further developments.

The capacity of certain channels can be approximated using tools from operator-valued free probability theory, an extension of Voiculescu's free probability theory \cite{DyNiVo92}. In this direction, Shlyakhtenko \cite{Sh96} showed how to compute the asymptotic eigenvalue distribution of band Gaussian matrices, i.e., the random matrices associated to Gaussian IND channels, using operator-valued free probability theory; from which the computation of the asymptotic capacity follows by a standard argument \cite{Te99}. These free probability approaches have also been applied in other contexts, such as frequency-selective slow-fading channels, where Rashidi Far et al. \cite{BrFaOrSp08} approximated the capacity of a channel with an associated block random matrix in which an arbitrary correlation between blocks was allowed. Their research is heavily based on the free probability tools in Helton et al. \cite{HeRaSp07}. Nonetheless, they also claim that their main result can be derived from Girko's techniques \cite{Gi95}. The computation of the capacity of some multiple access channels has also been approached by Girko's deterministic equivalents in Couillet et al.  \cite{CoDeSi11}, and by the free deterministic equivalents introduced in Speicher et al. \cite{SpVa12}.

Even though the qualitative facet of operator-valued free probability theory is still under development, recent techniques based on matricial fixed point equations have made it possible to numerically compute the asymptotic spectral distribution of a wide class of random matrix ensembles \cite{BeSpTrVa14,BeMaSp15}. A situation where these techniques are particularly useful is in the study of sums of products of independent block Gaussian matrices (see Section \ref{Subsection:CompoundChannels}). If we start with independent matrices like the ones considered in \cite{BrFaOrSp08} where an arbitrary correlation between blocks is allowed, then the entries of a sum of products of these Gaussian matrices are neither Gaussian, nor independent, nor identically distributed. A main feature of the operator-valued free probability approach resides in the systematic use of the concept of freeness introduced by Voiculescu \cite{Vo85}, which allows us to compute the asymptotic eigenvalue distribution of algebraic operations (sums and products) of certain independent block random matrices. These operator-valued free probability techniques might be of general interest to the information theory community. In particular, they offer a complementary approach to the deterministic equivalent method, which in some situations allows us to perform some numerical computations more efficiently.

In the present paper we consider point-to-point multiantenna channels with certain block distributional symmetries which do neither require the entries of the channel matrix to be Gaussian, nor independent, nor identically distributed. A main contribution of this paper is a capacity theorem for these channels that might be regarded as a generalization of Telatar's theorem \cite{Te99}, which reduces the numerical optimization domain in capacity computation. With this information theoretic result and some free probability arguments, we prove an asymptotic capacity theorem that, in addition to reducing the optimization domain, does not depend on the dimension of the channel matrix. This theorem allows us to apply techniques already known by the information theory community \cite{BrFaOrSp08,HeRaSp07} as well as the subordination techniques recently introduced by Belinschi et al. in \cite{BeSpTrVa14,BeMaSp15} to numerically compute the asymptotic capacity of the channels under consideration.

More precisely, consider a single user multiantenna channel under the general linear model
\begin{equation*}
y=Hx+n,
\end{equation*}
where $x$ is the $t$-dimensional input vector, $y$ is the $r$-dimen\-sional output vector, $H$ is the $r\times t$ matrix of propagation coefficients, and $n$ is an $r$-dimensional additive Gaussian noise with covariance matrix ${\rm I}_r$. Assume that $H$, $x$ and $n$ are statistically independent and that the average transmitting power is limited by some constant $P>0$. Throughout this paper we will assume that both the channel state information (CSI) at the receiver and the channel distribution information (CDI) at the transmitter are available. To keep the analysis tractable, we also assume that $r=t=dN$ for some positive integers $d$ and $N$. The general case, when $r\neq t$, can be analyzed following the ideas introduced by Benaych-Georges in \cite{Benaych2009}.

Let $C(H_{dN})$ denote the ergodic capacity of a channel whose associated $dN\times dN$ random matrix is $H_{dN}$. By definition, the ergodic capacity $C(H_{dN})$ is given by
\begin{equation}
\label{eq:DefCapacity}
C(H_{dN}) = \sup_{f_x} I(x;(y,H_{dN})),
\end{equation}
where $f_x$ ranges over all possible input distributions such that the power constraint is satisfied and $I(x;(y,H_{dN}))$ is the mutual information when $x$ is distributed as $f_x$. Let $\CvM{n}{\C}$ denote the set of covariance matrices $n\times n$ with complex coefficients. One of the first theorems in the field of multiantenna communications, which is implicitly proved in the groundbreaking work of Telatar \cite{Te99} (see also \cite{LoTuVe06}), states that the ergodic capacity $C(H_{dN})$ of a channel with associated random matrix $H_{dN}$ is given by
\begin{equation}
\label{eq:MultiantennaCapacity}
C(H_{dN})=\max_{\begin{smallmatrix}Q\in \CvM{dN}{\C}\\\Tr{Q}=1\end{smallmatrix}} \E{\log\det\left({\rm I}_{dN}+PH_{dN}QH_{dN}^*\right)},
\end{equation}
where the expectation is taken with respect to $H_{dN}$. Note that this result drastically reduces the optimization domain in \eqref{eq:DefCapacity}.

In general, the optimization in \eqref{eq:MultiantennaCapacity} has to be performed numerically. In this sense, capacity theorems reduce the optimization domain further when extra hypotheses are provided. A good example of this is a classical theorem of Telatar. Let $W_{dN}$ be a $dN\times dN$ random matrix with all its entries i.i.d. Rayleigh distributed. The essential symmetry property of this channel is that $W_{dN} \stackrel{\mathcal{L}}{=} W_{dN}U_{dN}$ for all $dN\times dN$ unitary matrices $U_{dN}$. The capacity of $W_{dN}$ is then given by
\begin{equation}
\label{eq:TheoremTelatar}
C(W_{dN}) = \E{\log\det\left({\rm I}_{dN} + \frac{P}{dN}W_{dN}W_{dN}^*\right)}.
\end{equation}
In terms of reducing the optimization domain, this theorem is the best we can hope for as the numerical optimization domain was reduced to a single point. Channel distribution symmetries that reduce the numerical optimization domain to a single point have been studied before, see for example Tulino et al. \cite{LoTuVe06}.

A trade-off between the numerical optimization problem and the generality of the channel distribution is evident. For example, Telatar's theorem does not require any numerical optimization at all, but the conditions imposed on the channel distribution might be restrictive. On the other hand, equation \eqref{eq:MultiantennaCapacity} holds for any random matrix distribution, but the numerical optimization domain has topological dimension $(dN)^2$, see \cite[Section 18.5]{Ga11}.

This paper explores a middle point in this trade-off. We assume weaker symmetry conditions in the channel distribution than in Telatar's canonical model. This allows us to encompass a wider range of applications at the expense of being able to prove merely that the capacity is achieved in a subset of $\CvM{dN}{\C}$ with topological dimension $d^2$. We also show how the aforementioned operator-valued free probability tools can be employed to compute the asymptotic capacity as $N\to\infty$ of certain channels with the considered symmetries. Altogether the optimum of the reduced optimization can be approximated using operator-valued free probability techniques.

In particular, we define $\class{S}_{d,N}$-channels to be channels with an associated $dN\times dN$ random matrix $H_{dN}$ such that
\begin{equation*}
H_{dN} \stackrel{\mathcal{L}}{=} H_{dN} ({\rm I}_d\otimes U_N)
\end{equation*}
for all $N\times N$ unitary matrices $U_N$. Particular examples of these channels are block Gaussian, compound and block Gaussian mixture channels (see Section \ref{Section:Applications}). Also we define $\class{B}_{d,N}$-channels as channels that have a capacity achieving input covariance matrix with a block structure $\frac{P}{N}\Phi\otimes{\rm I}_N$ where $\Phi\in\CvM{d}{\C}$. A main feature is that $\class{S}$-channels are also $\class{B}$-channels, i.e., we have found a sufficient condition for a channel to have a capacity achieving input covariance matrix with the aforementioned block structure. Indeed, this theorem can be regarded as a block-based generalization of Telatar's theorem.

For a general $\class{B}_{d,N}$-channel, the numerical optimization required to compute its capacity is essentially done over $\CvM{d}{\C}$, thus the optimization domain has dimension $d^2$. Another main result, in the spirit of the free deterministic equivalents introduced by Speicher et al. in \cite{SpVa12}, is that for certain $\class{B}$-channels it is possible to approximate both the capacity and a capacity achieving input covariance matrix using free probability tools. With this free probability approach, the corresponding optimization process is independent of $N$. Indeed, all the computations are performed using $2d\times 2d$ matrices. These two main results provide a very efficient method to approximate the capacity and capacity achieving input covariance matrices of certain $\class{S}_{d,N}$-channels.

Another important property of $\class{S}$-channels is that they posses a pseudo-algebraic structure. Namely, the sum of the products of independent matrices associated to $\class{S}_{d,N}$-channels is the matrix associated to an $\class{S}_{d,N}$-channel. In fact, compound channels are constructed in this way, e.g., $H_1H_2H_3H_4$, $H_1H_2H_3+H_4$ and $H_1H_2+H_3H_4$, where $\{H_1,\ldots,H_4\}$ are independent block Gaussian matrices like the ones in \cite{BrFaOrSp08}. The block Gaussian matrices in \cite{BrFaOrSp08}, which have an arbitrary correlation between blocks, are the random matrices associated to certain $\class{S}_{d,N}$-channels (see subsection \ref{Subsection:BlockGaussianChannelsRevisited}). The concatenated product of standard Gaussian matrices, which is a particular case of a compound channel, has been studied by Muller in \cite{Mu02} and references therein. Within the context of relay networks, Fawaz et al. \cite{DeFaGeZa11} analyzed the concatenated product of Kronecker correlated channels using free probability tools. Even though compound channels do not generalize the model in the latter reference, one of our main objectives is to provide a framework in which certain relay channels might be systematically studied. To the best of the authors's knowledge, the present paper is the first time that a general algorithm to deal with this kind of compound channels has been discussed.

The present paper is organized as follows. Section \ref{Section:MainResults} gives precise statements of the main results. Section \ref{Section:Applications} applies these results to some particular models: first we consider channels whose associated random matrices are block Gaussian matrices, and then we study compound channels, where the subordination techniques mentioned before play a central role. Section \ref{Section:Applications} ends with what we call the block Gaussian mixture channel. For the reader's convenience, Appendix \ref{Appendix:Notation} summarizes the notation used in this paper. Appendix \ref{Appendix:ProofMainResults} provides the proofs of the main results and Appendix \ref{Appendix:FreeProbability} summarizes the main concepts of free probability needed in this paper.

\section{Main results}
\label{Section:MainResults}

This section presents the main results of this paper. All proofs are deferred to Appendix \ref{Appendix:ProofMainResults}.

\subsection{$\class{B}$-channels and $\class{S}$-channels}

The two classes of channels considered in this paper are the $\class{B}$-channels and the $\class{S}$-channels. As mentioned before, \mbox{$\class{B}$-channels} have a capacity achieving input covariance matrix with a certain block structure, while $\class{S}$-channels have a distribution satisfying certain symmetry properties. In this way, $\class{B}$ stands for block and $\class{S}$ for symmetry.

\begin{definition}
For $d$ and $N$ positive integers, we define $\class{B}_{d,N}$ as the set of $dN\times dN$ random matrices $H_{dN}$ such that, for all $P>0$, its capacity $C(H_{dN})$ equals
\begin{equation}
\label{eq:CapacityBChannel}
\max_{\begin{smallmatrix}\Phi\in\CvM{d}{\C}\\\Tr{\Phi}=1\end{smallmatrix}} \E{\log\det\left({\rm I}_{dN}+\frac{P}{N}H_{dN}(\Phi\otimes{\rm I}_{N})H_{dN}^*\right)}.
\end{equation}
A channel is said to be a $\class{B}$-channel if its associated random matrix belongs to $\class{B}:=\bigcup_{d,N\in\N} \class{B}_{d,N}$.
\end{definition}

The previous definition is equivalent to saying that a capacity achieving input covariance matrix has the form $\frac{P}{N}\Phi\otimes{\rm I}_N$. However, this condition as stated in the previous definition allows us to see how to generalize the definition of capacity to certain operators. This is crucial in order to be able to compute the asymptotic capacity using free probability techniques.

The numerical optimization performed to obtain the capacity of a general multiantenna channel, as given in \eqref{eq:MultiantennaCapacity}, is done in a space of topological dimension $(dN)^2$. The definition of $\class{B}$-channels requires the same optimization, as given in \eqref{eq:CapacityBChannel}, to be done in a particular subspace of topological dimension $d^2$. Although the optimization is done in such a space, the evaluation of the expected value with respect to $H_{dN}$ still depends on $N$. For $N$ large, even a Monte Carlo based approach might be prohibitive, as the matrices involved can be very large. As we shall see in subsection \ref{Subsection:ClassB}, the capacity and capacity achieving input covariance matrix of certain \mbox{$\class{B}$-channels} can be approximated using free probability methods. These methods depend only on $d$ and not on $N$. In fact, the larger the $N$, the better the approximation.

The defining property of $\class{B}$-channels \eqref{eq:CapacityBChannel} is not straightforward to verify. In order for the definition of $\class{B}$-channels to be useful, we need symmetry conditions in the channel distribution that imply being a $\class{B}$-channel. This is the role of the concept of $\class{S}$-channels.

\begin{definition}
For $d$ and $N$ positive integers, we define $\class{S}_{d,N}$ as the set of $dN\times dN$ random matrices $H_{dN}$ such that for any $N\times N$ unitary matrix $U_N$,
\begin{equation}
\label{eq:DefiningConditionSChannels}
H_{dN} \stackrel{\mathcal{L}}{=} H_{dN}({\rm I}_d\otimes U_N).
\end{equation}
A channel is said to be an $\class{S}$-channel if its associated random matrix belongs to $\class{S}:=\bigcup_{d,N\in\N} \class{S}_{d,N}$.
\end{definition}

With respect to Telatar's model, where the channel matrix $H_{dN}$ satisfies the symmetry condition $H_{dN} \stackrel{\mathcal{L}}{=} H_{dN} U_{dN}$ for all $dN\times dN$ unitary matrix $U_{dN}\in\calU(dN)$, our model \eqref{eq:DefiningConditionSChannels} is less restricted, since ${\rm I}_d \otimes \calU(N) \subset \calU(dN)$. In fact,  condition \eqref{eq:DefiningConditionSChannels} does not require the blocks of $H_{dN}$ to be independent. For example, if $H_N^{(1)}$ and $H_N^{(2)}$ are two independent $N\times N$ Gaussian matrices as in the canonical model, then
\begin{equation*}
H_{2N} = \left(
\begin{matrix}
H_N^{(1)} & H_N^{(2)}\\
H_N^{(2)} & H_N^{(1)}
\end{matrix}
\right)
\end{equation*}
is an $\class{S}_{2,N}$-channel and its entries are clearly not independent. More examples will be consider in Section \ref{Section:Applications}.

\begin{remark}
\label{Remark}
Let $\pi$ be an element  of $\calS_N$, the symmetric group on $N$ letters. Abusing notation, we also denote by $\pi$ its associated permutation matrix, i.e., the $N\times N$ matrix defined by $\pi_{i,j} = \delta_{\pi(i)}(j)$ for all $1\leq i,j\leq N$. Let $\{e_1,\ldots,e_N\}$ be the standard basis in $\C^N$. Denote by $U_0$ the matrix whose columns are the vectors obtained by applying the Gram--Schmidt process to the basis
\begin{equation*}
\{e_1+\cdots+e_N,e_1-e_2,e_2-e_3,\ldots,e_{N-1}-e_N\}.
\end{equation*}
With this notation and in the spirit of \cite{LoTuVe06}, we can replace the condition defining $\class{S}$-channels by the following pair of weaker conditions,
\begin{itemize}
	\item[a)] $H_{dN}\stackrel{\mathcal{L}}{=}H_{dN}({\rm I}_d\otimes\pi)$ for all $\pi\in\calS_N$,
	\item[b)] $H_{dN}\stackrel{\mathcal{L}}{=}H_{dN}({\rm I}_d\otimes U_0)$,
\end{itemize}
and still be able to prove the main results concerning \mbox{$\class{S}$-channels}. A routine verification shows that the two conditions in question are equivalent to requiring that the distribution of $H_{dN}$ be invariant under right multiplication by the closure of the group generated by $\{{\rm I}_d\otimes U_0\}\cup({\rm I}_d\otimes \calS_N)$, which is clearly a subgroup of ${\rm I}_d\otimes \mathcal{O}(N)$ (here, $\mathcal{O}(N)$ is the orthogonal matrices of dimension $N$). In particular, if $H_{dN}$ is a $dN\times dN$ random matrix such that $H_{dN}\stackrel{\mathcal{L}}{=} H_{dN}({\rm I}_d\otimes O_N)$ for any orthogonal matrix $O_N\in\mathcal{O}(N)$, then $H_{dN}$ satisfies the previous conditions. From a modelling point of view, the condition in \eqref{eq:DefiningConditionSChannels} seems more natural, so we stick to that condition.
\end{remark}

The next theorem provides the aforementioned link between $\class{B}$-channels and $\class{S}$-channels. In particular, $\class{S}$-channels are also $\class{B}$-channels with the same block structure.

\begin{theorem}
\label{Theorem:ClassSImpliesClassB}
If $H_{dN}$ is a $dN\times dN$ random matrix in $\class{S}_{d,N}$, then $H_{dN}\in\class{B}_{d,N}$, i.e., the ergodic capacity $C(H_{dN})$ equals
\begin{equation*}
\max_{\begin{smallmatrix}\Phi\in\CvM{d}{\C}\\\Tr{\Phi}=1\end{smallmatrix}} \E{\log\det\left({\rm I}_{dN}+\frac{P}{N}H_{dN}(\Phi\otimes{\rm I}_{N})H_{dN}^*\right)}.
\end{equation*}
\end{theorem}

An important feature of the above theorem is that for \mbox{$\class{S}_{d,N}$-channels}, the capacity is achieved by a covariance matrix in $\CvM{d}{\C}\otimes{\rm I}_N \subset \CvM{dN}{\C}$, which reduces the dimension of the numerical optimization domain from $(dN)^2$ to $d^2$. In fact, the canonical model for multiantenna channels is an \mbox{$\class{S}_{1,N}$-channel}. In this way, Theorem~\ref{Theorem:ClassSImpliesClassB} is a generalization of Telatar's theorem \cite{Te99}.

Roughly speaking, the previous theorem says that the block symmetries satisfied by the distribution of an $\class{S}$-channel bequeath a block structure to a capacity achieving input covariance matrix, i.e., whenever $H_{dN} \stackrel{\mathcal{L}}{=} H_{dN}({\rm I}_d\otimes U_N)$ for all $U_N\in\calU(N)$, then there is a capacity achieving input covariance matrix with block form $\frac{P}{N}\Phi_0\otimes{\rm I}_N$ for some $\Phi_0\in\CvM{d}{\C}$. Observe that $\Phi_0$ is not necessarily the normalized identity matrix $\frac{1}{d}{\rm I}_d$ which would be the case in the $dN\times dN$ canonical model of Telatar \cite{Te99}. In fact, $\Phi_0$ depends on the correlation between the blocks of $H_{dN}$.

Another important feature of $\class{S}$-channels is the following algebraic-like property. This property allows us to construct more general $\class{S}$-channels from simpler ones, which provide us a more flexible family of models.

\begin{proposition}
\label{Proposition:AlgebraicClassS}
Let $d$ and $N$ be positive integers. Suppose that $H_1$ and $H_2$ are two independent random matrices in $\class{S}_{d,N}$. Then both $H_1+H_2$ and $H_1H_2$ belong to $\class{S}_{d,N}$.
\end{proposition}

In general, if $H_1$ and $H_2$ belong to $\class{S}_{d,N}$ then $H_1+H_2$ does not necessarily belong to $\class{S}_{d,N}$; similarly for $H_1H_2$. However, under the independence assumption in the previous proposition, $\class{S}_{d,N}$ has an algebraic-like behavior. This pseudo-algebraic behavior takes full advantage of the subordination methods seen in detail in Section \ref{Subsection:CompoundChannels}.

\subsection{Capacity and capacity achieving input covariance matrix of $\class{B}$-channels}
\label{Subsection:ClassB}

As said before, it is possible to approximate the capacity and a capacity achieving input covariance matrix of certain $\class{B}$-channels using free probability tools. In particular, there is an algorithm that relies on matricial fixed point equations which approximate both the capacity and a capacity achieving input covariance matrix using only $2d\times 2d$ matrices in the computation process.

In Appendix \ref{Appendix:FreeProbability} we summarize the main concepts of free probability theory used in this paper, and in the following section we show how to apply them to particular examples. Let $\Mat{d}{\calA}$ be the set of $d\times d$ matrices whose entries are non-commutative random variables in a non-commutative probability space $(\calA,\FI)$ and let $F^{\bf X}$ denote the spectral distribution of a selfadjoint element ${\bf X}$ in $\Mat{d}{\mathcal{A}}$.

We define the abstract capacity for elements in $\Mat{d}{\calA}$ as follows.

\begin{definition}
Let $P>0$ be fixed. The capacity functional $C:\Mat{d}{\calA}\to\R_+$ is defined for ${\bf H}\in\Mat{d}{\calA}$ by
\begin{equation*}
C({\bf H}) = \max_{\begin{smallmatrix}\Phi\in\CvM{d}{\C}\\\Tr{\Phi}=1\end{smallmatrix}} \int_{\R_+} \log(1+Px) \textrm{d} F^{{\bf H}\Phi{\bf H}^*}(x).
\end{equation*}
We say that $P\Phi_0$ is a capacity achieving input covariance matrix for the abstract capacity of ${\bf H}\in\Mat{d}{\calA}$ if
\begin{equation*}
C({\bf H}) = \int_{\R_+} \log(1+Px) \textrm{d} F^{{\bf H}\Phi_0{\bf H}^*}(x).
\end{equation*}
\end{definition}

In the applications of the following section we show how to deal with the numerical computation of $F^{{\bf H}\Phi{\bf H}^*}$. With this algorithm, we can compute the abstract capacity and an abstract capacity achieving input covariance matrix using any concave optimization method, since
\begin{equation*}
\Psi(\Phi) = \int_{\R_+} \log(1+Px)\textrm{d} F^{\mathbf{H}\Phi\mathbf{H}^{\ast }}(x)
\end{equation*}
is concave in $\Phi\in\CvM{d}{\C}$. To see the latter, observe that
\begin{equation*}
\Psi(\Phi)=\tau(\log(1+P{\bf H}\Phi{\bf H}^*)),
\end{equation*}
where $\tau:\Mat{d}{\calA}\to\C$ is the state $\TR\otimes\FI$, when we make the identification $\Mat{d}{\calA}\cong\Mat{d}{\C}\otimes\calA$. Since the logarithm is operator concave (see \cite{Fu92}, for example), the claim follows. It is important to remark that the algorithm to compute $F^{{\bf H}\Phi{\bf H}^*}$ uses $2d\times 2d$ matrices instead of $dN\times dN$ matrices as in a Monte Carlo scheme.

The following theorem provides us with a connection between the capacity in its abstract form and the capacity of $\class{B}$-channels.

\begin{theorem}
\label{Theorem:OpVCapacity}
Fix $d\in\N$. Suppose that $H_{dN}\in\class{B}_{d,N}$ for all $N\in\N$ and $(H_{dN}^{(i,j)}/\sqrt{N})_{i,j=1}^d$ converges in $*$-distribution to $({\bf H}_{i,j})_{i,j=1}^d$ for some ${\bf H}\in\Mat{d}{\calA}$. Then
\begin{equation*}
C(\mathbf{H)}=\lim_{N\rightarrow \infty }\frac{C(H_{dN})}{dN}.
\end{equation*}
Moreover, if $P\Phi_0$ is any capacity achieving input covariance matrix for the abstract capacity of ${\bf H}$, then 
\begin{equation*}
C(\mathbf{H)} = \lim_{N\to\infty} \frac{I(x^{\Phi_0,N};(y^{\Phi_0,N},H_{dN}))}{dN} 
\end{equation*}
where $I(x^{\Phi_0,N};(y^{\Phi_0,N},H_{dN}))$ is the mutual information of the system when the $dN$-dimensional input vector has covariance $\frac{P}{N}\Phi_0\otimes{\rm I}_{N}$.
\end{theorem}

This theorem shows that $C(\mathbf{H)}$ is the limit of the normalized capacities $C(H_{dN})$ and that $\frac{P}{N}\Phi_0\otimes{\rm I}_{N}$ is asymptotically optimal in the sense that
\begin{equation*}
\lim_{N\to\infty} \frac{C(H_{dN})-I(x^{\Phi_0,N};(y^{\Phi_0,N},H_{dN}))}{dN}=0.
\end{equation*}
That is, the difference between the normalized capacity and the normalized mutual information produced by $\frac{P}{N}\Phi_0\otimes{\rm I}_N$ goes to zero as $N\to\infty$. Thus we can approximate a capacity achieving input covariance matrix for the $dN\times dN$ system with $\frac{P}{N}\Phi_0\otimes{\rm I}_N$.

In order to apply this free probability based method to a random matrix $H_{dN}\in\class{B}_{d,N}$, it is necessary to associate to it an operator ${\bf H}\in\Mat{d}{\calA}$. Of course the associated operator depends on the specific distribution of $H_{dN}$. In general, this step is done invoking general theorems that ensure the convergence in $\ast$-distribution of certain ensembles of random matrices to non-commutative random variables. In the following section we illustrate this in a variety of examples.

\section{Applications}
\label{Section:Applications}

In this section we apply the theory developed in the preceding section to three families of channels: block Gaussian, compound, and block Gaussian mixture channels.

\subsection{Block Gaussian channels}
\label{Subsection:BlockGaussianChannelsRevisited}

In our setting, a block Gaussian channel is a multiantenna channel under the general linear model considered in the Introduction whose associated random matrix $H_{dN}$ is given by $H_{dN} = (H^{(i,j)})_{i,j=1}^d$ where $H^{(i,j)}$ for $1\leq i,j \leq d$ are $N\times N$ Gaussian matrices with the correlation structure
\begin{equation}
\label{eq:DefBlockGaussian}
\E{H_{r,p}^{(i,k)}\overline{H_{s,q}^{(j,l)}}} = \delta_{r,s} \delta_{p,q} \tau(i,k;j,l),
\end{equation}
for some covariance mapping $\tau(i,k;j,l)$ ($1\leq i,k,j,l\leq d$). Even though this channel is of interest in itself, its main role is to serve as the basic building block for the main application of this paper, the compound channels.

The random matrix model of the block Gaussian channels, which includes the random matrix model in \cite{Te99}, has been already used in Rashidi Far et al. \cite{BrFaOrSp08} in the context of frequency-selective slow-fading multiantenna channels. In that context, no numerical optimization was required. In our setting, this is no longer the case. Section \ref{Subsubsection:BlockS} will show that the block Gaussian channel is an $\class{S}$-channel, and so by our main theorems this allows us to approximate its capacity and a capacity achieving input covariance matrix using free probability tools in Section \ref{Subsubsection:AsymptoticCapacityBlock}. At the end of this subsection, we will make some modelling remarks concerning block Gaussian channels.

\subsubsection{A block Gaussian channel is an $\class{S}$-channel}
\label{Subsubsection:BlockS}

In what follows we provide an explicit construction for the random matrix model associated to a block Gaussian channel. This construction easily shows that it is an $\class{S}$-channel.

Let $\tau(i,k;j,l)$ for $1\leq i,j,k,l \leq d$ be a given covariance mapping and define $K$ as the $d^2 \times d^2$ complex covariance matrix given by $K_{i+(k-1)d,j+(l-1)d}=\tau(i,k;j,l)$. Let $W$ be a $d^2N \times N$ random matrix with i.i.d. circularly symmetric complex Gaussian entries with zero mean and unit variance. For $1\leq i\leq d^2$, write $W^{(i)}$ for the $N\times N$ random matrices satisfying $W=(W^{(1)T}\ W^{(2)T}\ \cdots\ W^{(d^2)T})^T$. We define $V=(K^{1/2}\otimes{\rm I}_N) W$ and for $1\leq i\leq d^2$ let $V^{(i)}$ be the $N\times N$ random matrices satisfying $V = (V^{(1)T}\ V^{(2)T}\ \cdots\ V^{(d^2)T})^T$. A straightforward computation shows that
\begin{equation}
\label{eq:VKW}
V^{(i)} = \sum_{k=1}^{d^2} (K^{1/2})_{i,k} W^{(k)},
\end{equation}
for all $1\leq i\leq d^2$. For $1\leq i,j\leq d$, define
\begin{equation*}
H^{(i,j)}:=V^{(i+(j-1)d)} \textnormal{ and } H_{dN}:=(H^{(i,j)})_{i,j=1}^d.
\end{equation*}
By the previous equation we have that ${\mathbf E}(H^{(i,k)}_{r,p}\overline{H^{(j,l)}_{s,q}})$ equals
\begin{equation*}
\sum_{a,b=1}^{d^2} (K^{1/2})_{i+(k-1)d,a} \overline{(K^{1/2})_{j+(l-1)d,b}} \E{W^{(a)}_{r,p} \overline{W^{(b)}_{s,q}}}.
\end{equation*}
Since all the entries of $W$ are independent, the latter expression equals
\begin{equation*}
\delta_{r,s}\delta_{p,q} \sum_{a=1}^{d^2} (K^{1/2})_{i+(k-1)d,a} (K^{1/2})_{a,j+(l-1)d}.
\end{equation*}
A straightforward manipulation leads to
\begin{equation*}
\E{H^{(i,k)}_{r,p}\overline{H^{(j,l)}_{s,q}}} = \delta_{r,s}\delta_{p,q} \tau(i,k;j,l).
\end{equation*}
Thus $H_{dN}$ as constructed above is the random matrix associated to the block Gaussian channel with covariance $\tau$.

Let $U_N$ be an $N\times N$ unitary matrix. Since $W^{(i)}\stackrel{\mathcal{L}}{=}W^{(i)}U_N$ for all $1\leq i\leq d^2$ and also $W^{(i)}U_N$ is independent from $W^{(j)}U_N$ for all $i\neq j$, then
\begin{equation*}
(W^{(1)}\ \cdots\ W^{(d^2)}) \stackrel{\mathcal{L}}{=} (W^{(1)}U_N\ \cdots\ W^{(d^2)}U_N),
\end{equation*}
i.e., $W\stackrel{\mathcal{L}}{=} WU_N$. Since $V=(K^{1/2}\otimes{\rm I}_N) W$, it follows that $V \stackrel{\mathcal{L}}{=} VU_N$. By definition of $H^{(i,j)}$ and $H_{dN}$ we have that
\begin{equation*}
H_{dN}({\rm I}_d \otimes U_N) = (H^{(i,j)}U_N)_{i,j=1}^d=(V^{i+(j-1)d}U_N)_{i,j=1}^d.
\end{equation*}
Therefore $H_{dN} \stackrel{\mathcal{L}}{=} H_{dN} ({\rm I}_d\otimes U_N)$, i.e., this block Gaussian channel is an $\class{S}$-channel.

\subsubsection{Asymptotic capacity of a block Gaussian channel}
\label{Subsubsection:AsymptoticCapacityBlock}

As shown in the previous subsubsection, a block Gaussian channel is an $\class{S}$-channel and thus, by Theorem~\ref{Theorem:ClassSImpliesClassB}, it is a $\class{B}$-channel. In order to approximate its capacity by means of Theorem~\ref{Theorem:OpVCapacity}, it is necessary to assign an operator ${\bf H}\in\Mat{d}{\calA}$ to the block Gaussian channel $H_{dN}=(H_N^{(i,j)})_{i,j=1}^d$. This is the objective of this subsubsection.

A theorem of Voiculescu (see Theorem~\ref{Thm:VoiculescuThm}) states that $(W_N^{(i)})_{1\leq i\leq d^2}$ converges in $*$-distribution as $N\to\infty$ to a free circular family $({\bf W}_i)_{1\leq i\leq d^2}\subset\calA$ in some non-commutative probability space $(\calA,\FI)$. Equation \eqref{eq:VKW} shows then that
\begin{equation*}
(H_N^{(i,j)})_{i,j=1}^d \stackrel{*\textnormal{-dist}}{\longrightarrow} ({\bf H}_{i,j})_{i,j=1}^d,
\end{equation*}
where
\begin{equation*}
{\bf H}_{i,j} = \sum_{k=1}^{d^2} (K^{1/2})_{i+(j-1)d,k} {\bf W}_k.
\end{equation*}
In particular, $({\bf H}_{i,j})_{i,j=1}^d$ is a circular family with covariance $\tau$. Indeed, $\F{H_{i,k}H_{j,l}^*}$ equals
\begin{equation*}
\sum_{a,b=1}^{d^2} (K^{1/2})_{i+(k-1)d,a} \overline{(K^{1/2})_{j+(l-1)d,b}} \F{{\bf W}_a{\bf W}_b^*}.
\end{equation*}
Since $({\bf W}_i)_{1\leq i\leq d^2}$ are free, we conclude that
\begin{align*}
\F{H_{i,k}H_{j,l}^*} &= \sum_{a=1}^{d^2} (K^{1/2})_{i+(k-1)d,a} (K^{1/2})_{a,j+(l-1)d}\\
&= K_{i+(k-1)d,j+(l-1)d}\\
&= \tau(i,k;j,l).
\end{align*}
This application of Theorem~\ref{Thm:VoiculescuThm} provides the operator-valued random variable ${\bf H}\in\Mat{d}{\calA}$ associated to $H_{dN}$.

By Theorem~\ref{Theorem:OpVCapacity}, we can approximate $C(H_{dN})$ by $dN C({\bf H})$ where
\begin{equation*}
C({\bf H}) = \sup_{\begin{smallmatrix}\Phi\in\CvM{d}{\C}\\\Tr{\Phi}=1\end{smallmatrix}} \int_\R \log(1+Px) \textrm{d} F^{{\bf H}\Phi{\bf H}^*}(x).
\end{equation*}
As mentioned in the previous section, the mapping
\begin{equation*}
\Phi\mapsto \int_\R \log(1+Px) \textrm{d} F^{{\bf H}\Phi{\bf H}^*}(x)
\end{equation*}
is concave in $\Phi$. Thus, if we are able to compute $F^{{\bf H}\Phi{\bf H}^*}$ for $\Phi\in\CvM{d}{\C}$, then we can apply any concave optimization method to numerically find $C({\bf H})$ and a covariance matrix that achieves it. Lets focus on computing $F^{{\bf H}\Phi{\bf H}^*}$.

Observe that ${\bf H}\Phi{\bf H}^*= ({\bf H}\Phi^{1/2})({\bf H}\Phi^{1/2})^*$, and write \mbox{${\bf X}={\bf H}\Phi^{1/2}$.} Since the entries of ${\bf X}$ are linear combinations of the entries of ${\bf H}$, they are also a circular family with covariance
\begin{align}
\nonumber \sigma(i,k;j,l) :=& \F{{\bf X}_{i,k}{\bf X}_{j,l}^*}\\
\nonumber =& \F{\left(\sum_{a=1}^d{\bf H}_{i,a}\Phi^{1/2}_{a,k}\right)\left(\sum_{b=1}^d{\bf H}_{j,b}\Phi^{1/2}_{b,l}\right)^*}\\
\label{eq:Sigma} =& \sum_{a,b=1}^d \Phi^{1/2}_{a,k} \overline{\Phi^{1/2}_{b,l}} \tau(i,a;j,b).
\end{align}
Recall that $\Mat{d}{\calA}$ can be identified with $\Mat{d}{\C}\otimes\calA$, thus ${\bf 1}_{\Mat{d}{\C}}\otimes\FI$ denotes the mapping from $\Mat{d}{\calA}$ to $\Mat{d}{\C}$ given by applying $\FI$ entrywise. Using $\sigma$ and the fixed point equation tools in Helton et al. \cite{HeRaSp07}, one can numerically compute the $\Mat{2d}{\C}$-valued Cauchy transform $\calG_{\widehat{\bf X}}$ of
\begin{equation*}
\widehat{\bf X} = \left(\begin{matrix}0 & {\bf X}\\{\bf X}^* & 0\end{matrix}\right).
\end{equation*}
In fact,  for $B\in\Mat{2d}{\C}^+$,
\begin{align}
\nonumber \calG_{\widehat{\bf X}}(B) :=& ({\bf 1}_{\Mat{d}{\C}}\otimes\FI)((\widehat{\bf X}-B)^{-1})\\
\label{eq:OpVCauchyBlock} =& \lim_{n\to\infty} T_B^{\circ n}(W)
\end{align}
for any $W\in\Mat{2d}{\C}^-$ where $T_B(W):=(B - \eta(W))^{-1}$ and $\eta(W):=({\bf 1}_{\Mat{d}{\C}}\otimes\FI)(\widehat{\bf X}W\widehat{\bf X})$. It is important to note that $\eta$ can be obtained directly from $\sigma$, see \cite{HeRaSp07}. Equation~\eqref{eq:OpVCauchyBlock} is easily implementable and does not depend on $N$. In fact, all the computations are done using $2d\times 2d$ matrices.

The spectral distribution of ${\bf X}{\bf X}^*={\bf H}Q{\bf H}^*$ can be obtained from the $\Mat{2d}{\C}$-valued Cauchy transform $\calG_{\widehat{\bf X}}$ as in \cite{BrFaOrSp08}. Specifically, the density\footnote{In order to have a spectral density, some regularity conditions on the spectral distribution are required. In most applied situations this is likely to be the case, as suggested by \cite{MaSpWe15}, \cite{Sh14}, and references therein.} $f^{{\bf XX}^*}$ of $F^{{\bf XX}^*}$ is given by
\begin{equation}
\label{eq:StieltjesInversionXX}
f^{{\bf XX}^*}(x) = -\frac{1}{\pi} \lim_{\epsilon\to0^+} \frac{\tr{\calG_{\widehat{\bf X}}(\sqrt{x+i\epsilon}{\rm I}_{2d})}}{\sqrt{x+i\epsilon}}.
\end{equation}

\subsubsection{Modelling features}

Let $\Phi:\Mat{dN}{\C}\to\Mat{dN}{\C}$ be the mapping given by
\begin{equation}
\label{eq:DefPhi}
\Phi(A) = \left(\begin{matrix}
(A_{1,1}^{(i,j)})_{i,j=1}^d & \cdots & (A_{1,N}^{(i,j)})_{i,j=1}^d\\
\vdots & \ddots & \vdots\\
(A_{N,1}^{(i,j)})_{i,j=1}^d & \cdots & (A_{N,N}^{(i,j)})_{i,j=1}^d
\end{matrix}\right),
\end{equation}
where $A=(A^{(i,j)})_{i,j=1}^d$ with $A^{(i,j)}\in\Mat{N}{\C}$ for all $1\leq i,j\leq d$. It is easy to see that $\Phi(A)$ is obtained by permuting the columns and rows of $A$. More specifically, we have that $\Phi(A)=\gamma A \gamma^{-1}$ for some $\gamma\in\calS_{dN}\subset\calU(dN)$ and in particular, the spectral distributions of $A$ and $\Phi(A)$ are the same. Thus all the theory developed so far applies to $\Phi(H_{dN})$ in the same way as it does to $H_{dN}$ after the appropriate adaptation by means of $\Phi$.

By \eqref{eq:DefPhi} and the definition of block Gaussian channel \eqref{eq:DefBlockGaussian}, we can see that $\Phi(H_{dN})$ is an $N\times N$ block random matrix with blocks of size $d\times d$ such that: a) the entries of each block have correlation $\tau$ and b) the different blocks are independent. Property a) might be interpreted as a uniform local behavior in which all the $d\times d$ blocks have the same distribution. Property b) suggests a short range dependency in which entries far enough away from each other become independent.

As a final remark, having $d\times d$ blocks of size $N\times N$ allows us to have the notion of convergence in $\ast$-distribution. This convergence is a key step in order to apply free probability techniques, thus we stick to it throughout this paper.

\subsection{Compound channels}
\label{Subsection:CompoundChannels}

In this subsection, we consider our main application: compound channels. A compound channel is a channel whose associated random matrix is the sum of products of independent block Gaussian matrices, that is, its associated random matrix has the form
\begin{equation}
\label{eq:DefCompoundChannels}
H_{dN} = \sum_{n=1}^s \prod_{m=1}^{p_n} H(n,m),
\end{equation}
for some $s\in\N$, $p_1,\ldots,p_s\in\N$ and independent $dN\times dN$ block Gaussian matrices
\begin{equation*}
\{H(n,m) : 1\leq n\leq s,1\leq m \leq p_n\}.
\end{equation*}
Observe that every block Gaussian matrix $H(n,m)$ has its own covariance mapping $\tau_{n,m}$. Of course $H(n,m)$ depends on $d$ and $N$, but we suppress this dependence to make the notation less cumbersome.

\subsubsection{Compound channels are $\class{S}$-channels}

Let $H_{dN}$ be a random matrix as in \eqref{eq:DefCompoundChannels}. By induction and Proposition~\ref{Proposition:AlgebraicClassS}, the random matrices $\{\prod_{m=1}^{p_n} H(n,m) : 1\leq n\leq s\}$ belong to $\class{S}_{d,N}$ and, since all the block Gaussian matrices are independent, they are independent. Again from induction and Proposition~\ref{Proposition:AlgebraicClassS}, we conclude that
\begin{equation*}
H_{dN}=\sum_{n=1}^s \prod_{m=1}^{p_n} H(n,m)
\end{equation*}
belongs to $\class{S}_{d,N}$. Therefore compound channels are also \mbox{$\class{S}$-channels} and in particular Theorem~\ref{Theorem:ClassSImpliesClassB} applies to this family of channels.

\subsubsection{Asymptotic capacity of compound channels}

To apply Theorem~\ref{Theorem:OpVCapacity}, the blocks of $H_{dN}$ have to converge in \mbox{$\ast$-distribution}, i.e., $\displaystyle (H_{dN}^{(i,j)})_{i,j=1}^d \stackrel{\ast\textnormal{-dist}}{\longrightarrow} ({\bf H}_{i,j})_{i,j=1}^d$ for some $\{{\bf H}_{i,j} : 1\leq i,j\leq d\} \subset \calA$. By the previous subsection, each block Gaussian matrix $H(n,m)$ has an associated ${\bf H}(n,m)$ in $\Mat{d}{\calA}$. Thus, it is natural to take
\begin{equation*}
{\bf H}=\sum_{n=1}^s \prod_{m=1}^{p_n} {\bf H}(n,m)
\end{equation*}
as the operator-valued random variable associated to $H_{dN}$. However, in order for this to work, we need to know the joint $\ast$-distribution of the families
\begin{equation*}
(\{{\bf H}(n,m)_{i,j} : 1\leq i,j\leq d\} : 1\leq n\leq s,1\leq m\leq p_n).
\end{equation*}
In what follows, we show that these families are free in $(\calA,\FI)$.

From the construction described in the previous section, for each $1\leq n\leq s$ and $1\leq m\leq p_n$, the blocks of $H(n,m)$ are linear combinations of some independent $N\times N$ standard Gaussian matrices, let's say $\{W(n,m)^{(i)} : 1\leq i\leq d^2\}$. Since $\{H(n,m) : n,m\}$ are independent, we may assume that all the random matrices $\{W(n,m)^{(i)} : n,m,i\}$ are independent. By Theorem~\ref{Thm:VoiculescuThm}, the family
\begin{equation*}
(W(n,m)^{(i)} : 1\leq n\leq s,1\leq m\leq p_n,1\leq i\leq d^2)
\end{equation*}
converges in $\ast$-distribution to a free circular family
\begin{equation*}
({\bf W}(n,m)^{(i)} : 1\leq n\leq s,1\leq m\leq p_n,1\leq i\leq d^2).
\end{equation*}
Since the entries of ${\bf H}(n,m)$ are linear combinations of $\{{\bf W}(n,m)^{(i)} : i\}$, the families $(\{{\bf H}(n,m)_{i,j} : i,j\}: n,m)$ are also free. Note that, for a given $n$ and $m$, the non-commutative random variables $\{{\bf H}(n,m)_{i,j} : i,j\}$ are not necessarily free.

Recall that the computation of the capacity functional $C({\bf H})$ of ${\bf H}\in\Mat{d}{\calA}$ requires computing the spectral distribution of ${\bf H}\Phi{\bf H}^*$ for $\Phi\in\CvM{d}{\C}$. However, as we saw in the previous subsection, the covariance structure in $\Phi$ can be melded with the covariance structures $\tau_{n,p_n}$ of the ${\bf H}(n,p_n)$, for each $1\leq n\leq s$, to form a new operator-valued random variable whose entries are a circular family with covariance $\sigma_{n,p_n}$, see \eqref{eq:Sigma}. Therefore, without loss of generality, we can focus on computing the operator-valued Cauchy transform of $\displaystyle \widehat{\bf H}=\left(\begin{smallmatrix} 0 & {\bf H}\\{\bf H}^* & 0\end{smallmatrix}\right)$, from which the spectral distribution is obtained as in \eqref{eq:StieltjesInversionXX}.

Let ${\bf A},{\bf B}\in\Mat{d}{\calA}$. An easy, yet crucial, fact states that if
\begin{equation*}
\{{\bf A}_{i,j} : 1\leq i,j\leq d\} \textnormal{ and }\{{\bf B}_{i,j} : 1\leq i,j\leq d\}
\end{equation*}
are free families, then ${\bf A}$ and ${\bf B}$ are in a relation called operator-valued freeness\footnote{More specifically, they are free with amalgamation over $\Mat{d}{\C}$.}. An in-depth discussion about this type of freeness is outside the scope of this paper, we refer the interested reader to \cite{Sp98} and references therein. Since the families $(\{{\bf H}(n,m)_{i,j} : i,j\}: n,m)$ are free, we have that $\{\widehat{{\bf H}(n,m)} : n,m\}$ are operator-valued free. The relevance of operator-valued freeness comes from the fact that it is needed to apply the subordination techniques in \cite{BeSpTrVa14,BeMaSp15}. In what follows we show how to use these techniques in the context of the present paper.

Consider the following variations of the Cauchy transform
\begin{align*}
r_{\bf X}(B) &= \calG_{\bf X}(B)^{-1}-B,\\
h_{\bf X}(B) &= B^{-1}-\calG_{\bf X}(B^{-1})^{-1}.
\end{align*}
Assume that ${\bf H}_1$ and ${\bf H}_2$ are two elements in $\Mat{d}{\calA}$ such that $\widehat{{\bf H}_1}$ and $\widehat{{\bf H}_2}$ are operator-valued free. Furthermore, assume that both $\calG_{\widehat{{\bf H}_1}}$ and $\calG_{\widehat{{\bf H}_2}}$ are given. Note that
\begin{equation*}
\widehat{{\bf H}_1+{\bf H}_2} = \widehat{{\bf H}_1} + \widehat{{\bf H}_2}.
\end{equation*}
By \cite{BeMaSp15}, for $B\in\Mat{2d}{\C}^+$,
\begin{equation}
\label{eq:OpVSum}
\calG_{\widehat{{\bf H}_1+{\bf H}_2}}(B) = \calG_{\widehat{{\bf H}_1}}(\omega_1(B)),
\end{equation}
where
\begin{equation*}
\omega_1(B)=\lim_{n\to\infty} f_B^{\circ n}(W)
\end{equation*}
for any $W\in\Mat{2d}{\C}^+$ and $f_B(W) = r_{\widehat{{\bf H}_2}}(r_{\widehat{{\bf H}_1}}(W)+B)+B$. Let $\widetilde{\bf X}=\left(\begin{matrix}{\bf X}^*{\bf X} & 0 \\ 0 & {\rm I}\end{matrix}\right)$. Since the state $\FI$ is tracial,
\begin{align*}
\widehat{{\bf H}_1{\bf H}_2} 
&= \left(\begin{matrix}{\bf H}_1 & 0\\ 0 & {\rm I}_d\end{matrix}\right)
\left(\begin{matrix} 0 & {\bf H}_2\\{\bf H}_2^* & 0\end{matrix}\right)
\left(\begin{matrix} {\bf H}_1^* & 0\\ 0 & {\rm I}_d\end{matrix}\right)\\
&\stackrel{\mathcal{L}}{=} \widetilde{{\bf H}_1}^{1/2} \widehat{{\bf H}_2} \widetilde{{\bf H}_1}^{1/2}.
\end{align*}
Let
\begin{equation*}
\omega_2(B)=\lim_{n\to\infty} g_B^{\circ n}(W)
\end{equation*}
for any $W\in\Mat{2d}{\C}^+$ and $g_b(W)=Bh_{\widetilde{{\bf H}_1}}(h_{\widehat{{\bf H}_2}}(W)B)$. By \cite{BeSpTrVa14}, for $B\in\Mat{2d}{\C}^+$,
\begin{equation}
\label{eq:OpVProduct}
\calG_{\widetilde{{\bf H}_1}\widehat{{\bf H}_2}}(B) = \calG_{\widehat{{\bf H}_2}}(\omega_2(B^{-1})^{-1})\omega_2(B^{-1})^{-1}B^{-1}.
\end{equation}
Note that in order to evaluate the previous equation, it is necessary to obtain first $\calG_{\widetilde{{\bf H}_1}}$. Applying the Schur complement formula, it can be verified that the $\Mat{2d}{\C}$-valued Cauchy transform of $\widetilde{{\bf H}_1}$ is given by \eqref{eq:tildeH1}, where
\begin{equation*}
g=\calG_{{\bf H}_1^*{\bf H}_1}(B_1+B_2({\rm I}_d-B_4)^{-1}B_3).
\end{equation*}
If the odd $\ast$-moments of $\{({\bf H}_1)_{i,j} : i,j\}$ vanish, i.e., 
\begin{equation*}
\F{x_1\ldots,x_{2n+1}}=0
\end{equation*}
for all $n\in\N$ and $x_1,\ldots,x_{2n+1}\in\{({\bf H}_1)_{i,j},({\bf H}_1)^*_{i,j} : i,j\}$, then
\begin{equation*}
\calG_{\widehat{{\bf H}_1}}\left(\begin{smallmatrix} {\rm I}_d & 0\\0&B\end{smallmatrix}\right) = \left(\begin{matrix}\ast & \ast\\ \ast & \calG_{{\bf H}_1^*{\bf H}_1}(B)\end{matrix}\right),
\end{equation*}
where the asterisks denote $d\times d$ complex matrices.

Recall that the elements $\{\widehat{{\bf H}(n,m)} : n,m\}$ are operator-valued free and their $\Mat{2d}{\C}$-valued Cauchy transforms are given by \eqref{eq:OpVCauchyBlock}. A routine computation shows that, for all $n$ and $m$, the odd \mbox{$\ast$-moments} of $\{{\bf  H}(n,m)_{i,j} : i,j\}$ vanish. These observations allow us to compute the $\Mat{2d}{\C}$-valued Cauchy transform of $\widehat{\bf H}$ by successive computation of the corresponding transforms of sums and products of pairs of operator-valued random variables, as in the previous paragraph. This systematic reduction is one of the main strengths of the free probability approach, and relies strongly on the notion of freeness. In this sense, when convergence in $\ast$-distribution is available, the free probability approach is a useful alternative to deterministic equivalent approaches.

An alternative approach to the successive computation described above comes from the linearization technique discussed in \cite{Sp15} and references therein. In the context of this paper, this approach changes the analysis of a sum of products of elements in $\Mat{d}{\calA}$ to the analysis of a single element in $\Mat{D}{\calA}$ for some $D\geq 2d$. Moreover, the $\Mat{D}{\C}$-valued Cauchy transform of such an element can be computed from \eqref{eq:OpVCauchyBlock} without using either \eqref{eq:OpVSum} or \eqref{eq:OpVProduct}. However, $D$ is usually much bigger that $2d$, hence loosing the sought dimensionality reduction. Further investigations are needed in order to determine which approach works better in concrete escenarios.

\subsubsection{A further application}
\label{Subsubsection:FurtherApplication}

Even though we defined compound channels to be channels whose associated random matrices are the sums of products of independent block Gaussian matrices, the techniques in this subsection applies in a more general setting. Let $H_{dN}=\sum_{n=1}^s \prod_{m=1}^{p_n} H(n,m)$ where the $H(n,m)$ are $dN\times dN$ random matrices, and let
\begin{equation*}
{\bf H}=\sum_{n=1}^s \prod_{m=1}^{p_n} {\bf H}(n,m),
\end{equation*}
where the $\{{\bf H}(n,m): n,m\}\subset\Mat{d}{\calA}$ are the operator-valued random variables associated to $\{H(n,m) : n,m\}$. If a) $H_{dN}$ is a $\class{B}$-channel, b) the elements $\{\widehat{{\bf H}(n,m)}: n,m\}$ are operator-valued free, and c) $\calG_{\widehat{{\bf H}(n,m)}}$ and $\calG_{\widetilde{{\bf H}(n,m)}}$ can be computed either explicitly or numerically, then we can replicate the analysis in the previous subsection. Observe that $H(n,m)$, $1\leq n\leq s$ and $1\leq m\leq p_n$, is not required to be a block Gaussian channel, which allows us to consider a wider range of applications. To illustrate this, consider the following example.

Let $H_{dN}$ be as in \eqref{eq:HdNModel}, where $W_{dN}$ is the random matrix associated to a compound channel, $\{U_N^{(i)},V_N^{(i)} : 1\leq i\leq d\}$ are independent $N\times N$ Haar unitary random matrices which are also independent from $W_{dN}$, and $\{A_N^{(i)},B_N^{(i)} : 1\leq i\leq d\}$ are selfadjoint constant matrices. It is straightforward to see that $H_{dN} ({\rm I}_d\otimes U_N) \stackrel{\mathcal{L}}{=} H_{dN}$ for all $U_N\in\calU(N)$, i.e., $H_{dN}$ is an $\class{S}$-channel and by Theorem~\ref{Theorem:ClassSImpliesClassB} it is also a $\class{B}$-channel.

\begin{figure*}[!t]
\normalsize
\begin{equation}
\label{eq:tildeH1}
\calG_{\widetilde{{\bf H}_1}}\left(\begin{smallmatrix} B_1 & B_2\\B_3 & B_4\end{smallmatrix}\right) = \left(\begin{matrix} g & g B_2 ({\rm I}_d-B_4)^{-1}\\({\rm I}_d-B_4)^{-1}B_3g & (B_4-{\rm I}_d)^{-1} + ({\rm I}_d-B_4)^{-1}B_3gB_2({\rm I}_d-B_4)^{-1} \end{matrix}\right)
\end{equation}
\begin{equation}
\label{eq:HdNModel}
H_{dN} = \left(\begin{smallmatrix} U_N^{(1)} A_N^{(1)} U_N^{(1)*} &  & \\ & \ddots & \\ & & U_N^{(d)}A_N^{(d)}U_N^{(d)*} \end{smallmatrix}\right)
W_{dN}
\left(\begin{smallmatrix} V_N^{(1)} B_N^{(1)} V_N^{(1)*} &  & \\ & \ddots & \\ & & V_N^{(d)}B_N^{(d)}V_N^{(d)*} \end{smallmatrix}\right)
\end{equation}
\hrulefill
\end{figure*}

Assume that each $A_N^{(i)}$ and $B_N^{(i)}$ for $1\leq i\leq d$ converge in distribution. Recall that the blocks of $W_{dN}$ converge in $\ast$-distribution to some, not necessarily circular, family $({\bf W}_{i,j})_{i,j=1}^d\subset\calA$. By Theorem~\ref{Thm:VoiculescuThm2},
\begin{align*}
& (U_N^{(1)}A_N^{(1)}U_N^{(1)*},\ldots,V_N^{(d)}B_N^{(d)}V_N^{(d)*},(W_{dN}^{(i,j)})_{i,j})
\end{align*}
converges in $\ast$-distribution to
\begin{equation*}
(a_1,\ldots,a_d,b_1,\ldots,b_d,({\bf W}_{i,j})_{i,j}),
\end{equation*}
where the non-commutative random variables $a_1,\ldots,a_d,$ $b_1,\ldots,b_d$ are selfadjoint, free among themselves, and free from $({\bf W}_{i,j})_{i,j}$. Thus, the operator-valued random variable ${\bf H}$ associated to $H_{dN}$ has the form
\begin{equation*}
{\bf H} = \left(\begin{smallmatrix}a_1 & &\\&\ddots &\\& & a_d\end{smallmatrix}\right) {\bf W} \left(\begin{smallmatrix}b_1 & &\\&\ddots &\\& & b_d\end{smallmatrix}\right)=:{\bf A}{\bf W}{\bf B}.
\end{equation*}

Since the entries of ${\bf A}$, ${\bf B}$ and ${\bf W}$ are free, we can use the subordination techniques previously discussed to compute the $\Mat{2d}{\C}$-valued Cauchy transform of
\begin{equation*}
\widehat{\bf H} = \left(\begin{matrix} {\bf A} & 0\\0&{\bf B}\end{matrix}\right) \widehat{\bf W}  \left(\begin{matrix} {\bf A} & 0\\0&{\bf B}\end{matrix}\right).
\end{equation*}
Note that the $\Mat{2d}{\C}$-valued Cauchy transform of $\widehat{\bf W}$ can be obtained as in the previous subsubsection. Thus, in order to compute $\calG_{\widehat{\bf H}}$, it is only necessary to find the $\Mat{2d}{\C}$-valued Cauchy transform of
\begin{equation*}
\left(\begin{matrix}{\bf A} & 0\\0&{\bf B}\end{matrix}\right)^2 = \sum_{k=1}^d a_k^2 E_{k,k} + \sum_{k=1}^d b_k^2 E_{d+k,d+k},
\end{equation*}
where $E_{k,k}$ denotes the $k,k$-unit matrix. Since the non-commutative random variables $a_1^2,\ldots,a_d^2,b_1^2,\ldots,b_d^2$ are free among themselves, additive subordination techniques reduce the problem to finding the $\Mat{2d}{\C}$-valued Cauchy transforms of the summands in the previous sum. In fact, if $x\in\calA$ is a non-commutative random variable and $E_{k,k}\in\Mat{2d}{\C}$ is the $k,k$-unit matrix, then $\calG_{aE_{k,k}}(B)$ equals, for $B\in\Mat{2d}{\C}^+$,
\begin{equation*}
B^{-1}+[B^{-1}]_{k,k}^{-2}\left(G_{x}([B^{-1}]_{k,k}^{-1})-[B^{-1}]_{k,k}\right) B^{-1}E_{k,k}B^{-1}.
\end{equation*}

As a final remark, every non-commutative random variable is free from constants. Moreover, the operator-valued Cauchy transform $\calG_M$ of a constant matrix $M$ is given by
\begin{equation*}
\calG_M(B)=(B-M)^{-1}.
\end{equation*}
Therefore, the operator-valued Cauchy transform of ${\bf H}\Phi{\bf H}^*$ for $\Phi\in\CvM{d}{\C}$ can be computed from $\calG_{\widehat{\bf H}}$ and $\calG_\Phi$ using the multiplicative subordination techniques already discussed.

\subsection{Block Gaussian mixture channels}
\label{Subsection:BlockMixtureChannels}

We now consider an extension of the concept of a block Gaussian channel, which is also an $\class{S}$-channel. The random matrix associated to such a channel, which might be regarded as a mixture of block Gaussian matrices, cannot be analyzed with the free probability tools of the two previous subsections. Theorem~\ref{Theorem:ClassSImpliesClassB} still applies since this channel is an $\class{S}$-channel; however, there is no notion of convergence in distribution and thus Theorem~\ref{Theorem:OpVCapacity} does not apply. The objective in introducing this channel is to show that the family of $\class{S}$-channels is richer than the family of channels for which Theorem~\ref{Theorem:OpVCapacity} applies. This reveals the flexibility of $\class{S}$-channels and suggests that many interesting applications are still to be discovered.

A mixture block Gaussian channel is a channel whose associated random matrix $H_{dN}$ has the form
\begin{equation*}
H_{dN} = (A \otimes {\rm J}_N) \circ X_{dN},
\end{equation*}
where $X_{dN}$ is block Gaussian matrix with covariance mapping $\tau$ and $A$ is a $d\times d$ random matrix independent from $X_{dN}$. Here ${\rm J}_N$ denotes the $N\times N$ matrix with all its entries equal to one and $\circ$ denotes the Hadamard or entrywise product.

Since ${\rm J}_N$ is constant, we have that $(A \otimes {\rm J}_N)$ and $X_{dN}$ are independent. Let $U_N\in\calU(N)$. Since $U_N$ is constant, the matrices $(A \otimes {\rm J}_N)$ and $X_{dN} ({\rm I}_d\otimes U_N)$ are also independent. The matrix $X_{dN}$ belongs to $\class{S}_{d,N}$, so we have the equality in distribution of $X_{dN} \stackrel{\mathcal{L}}{=} X_{dN} ({\rm I}_d\otimes U_N)$. Therefore
\begin{equation*}
H_{dN} \stackrel{\mathcal{L}}{=} H_{dN} ({\rm I}_d\otimes U_N),
\end{equation*}
i.e., the block Gaussian mixture channel is an $\class{S}$-channel. This implies that Theorem~\ref{Theorem:ClassSImpliesClassB} can be applied and thus the numerical optimization to compute the capacity is performed over a subset of $\CvM{d}{\C}\otimes{\rm I}_N$.

As in subsection \ref{Subsection:BlockGaussianChannelsRevisited}, the blocks of $X_{dN}$ converge in $\ast$-distribution to the entries of an operator-valued random variable ${\bf X}\in\Mat{d}{\calA}$. However, since $A$ is a $d\times d$ random matrix, there is no straightforward meaning for its limit in distribution as $N\to\infty$. This makes impossible the application of Theorem~\ref{Theorem:OpVCapacity} to approximate the capacity, and in particular we cannot neglect the dependency on $N$ in the final numerical optimization. The development of appropriate tools to deal with this kind of random matrices is postponed to future research.

\section*{Acknowledgements}

Mario Diaz was supported by the government of Ontario, Canada. The authors would like to thank James Mingo for valuable discussions while preparing this paper. This research was also supported by the Consejo Nacional de Ciencia y Tecnolog\'{i}a grant 222668, Mexico.

\appendices

\section{Notation}
\label{Appendix:Notation}

\begin{itemize}

\item[--] $\N$: the set of natural numbers;

\item[--] $\R$ and $\R_+$: the set of real and positive real numbers;

\item[--] $\C$: the set of complex numbers;

\item[--] $\Mat{n\times m}{\calA}$: the set of all $n\times m$ matrices with entries from the algebra $\calA$;

\item[--] $\Mat{n}{\calA}$: the set of all $n\times n$ matrices with entries from the algebra $\calA$;

\item[--] $\Re(a)=\frac{a+a^*}{2}$ and $\Im(a)=\frac{a-a^*}{2i}$: the real and imaginary part of an element $a$ in a $\ast$-algebra;

\item[--] $\Mat{n}{\C}^+$: the set of $n\times n$ matrices with positive definite imaginary part;

\item[--] $\Mat{n}{\C}^-$: the set of $n\times n$ matrices with negative definite imaginary part;

\item[--] $\CvM{n}{\C}$: $n\times n$ covariance matrices over the complex numbers;

\item[--] $\mathcal{O}(n)$: the set of orthogonal $n\times n$ matrices;

\item[--] $\calU(n)$: the set of unitary $n\times n$ matrices;

\item[--] $A_{i,j}$ or $(A)_{i,j}$: the $i,j$th entry of $A$;

\item[--] $E_{i,j}$: the $i,j$-unit matrix, i.e., the matrix with all its entries equal to zero except for the $i,j$th entry which equals one;

\item[--] $\text{Tr}(A)$: unnormalized trace of the matrix $A$;

\item[--] $\text{tr}\left(A\right)$: normalized trace of the matrix $A$;

\item[--] $A^{T}$: the transpose of the matrix $A$;

\item[--] $A^{\ast }$: the conjugate transpose of the matrix $A$;

\item[--] $\diag{a_1,\ldots,a_n}$: the $n\times n$ diagonal matrix with diagonal entries $a_1,\ldots,a_n$;

\item[--] $\widehat{A}$: the block matrix $\displaystyle \left(\begin{smallmatrix}0&A\\A^*&0\end{smallmatrix}\right)$;

\item[--] $\widetilde{A}$: the block matrix $\displaystyle \left(\begin{smallmatrix}A^*A&0\\0&{\rm I}\end{smallmatrix}\right)$;

\item[--] ${\rm I}_n$: the $n\times n$ identity matrix;

\item[--] ${\rm J}_n$: the $n\times n$ matrix with all its entries equal to one;

\item[--] $A\otimes B$: tensor product of two matrices, see \eqref{eq:TensorProductDef};

\item[--] $A\circ B$: Hadamard or entrywise product;

\item[--] ${\mathbf E}$: expected valued with respect to a classical probability space $(\Omega ,\mathcal{F},{\mathbf P})$;

\item[--] $F^X$: mean eigenvalue distribution of a random matrix $X$;

\item[--] $\calS_n$: the symmetric group on $n$ letters;

\item[--] $f^{\circ n}$: the $n$th composition of the function $f$;

\item[--] $F_n\Rightarrow G$: weak convergence of the distributions $(F_n)_{n=1}^\infty$ to $G$.

\end{itemize}

\section{Proofs of the main results}
\label{Appendix:ProofMainResults}

Before proving Theorem~\ref{Theorem:ClassSImpliesClassB}, we need to establish some facts of general interest.

For $A\in\Mat{p\times q}{\C}$ and $B\in\Mat{r\times s}{\C}$, we consider their tensor product $A\otimes B\in\Mat{pr\times qs}{\C}$,
\begin{equation}
\label{eq:TensorProductDef}
A\otimes B = \left(
\begin{matrix}
A_{1,1} B & A_{1,2} B & \cdots & A_{1,q} B\\
A_{2,1} B & A_{2,2} B & \cdots & A_{2,q} B\\
\vdots & \vdots & \ddots & \vdots\\
A_{p,1} B & A_{p,2} B & \cdots & A_{p,q} B\\
\end{matrix}
\right).
\end{equation}
Let $\pi$ be an element  of $\calS_N$, the symmetric group on $N$ letters. Abusing notation, we also denote by $\pi$ its associated permutation matrix, i.e., the $N\times N$ matrix defined by $\pi_{i,j} = \delta_{\pi(i)}(j)$ for all $1\leq i,j\leq N$. Recall that the $\pi^{-1} = \pi^T$.

\begin{lemma}
Let $\mathfrak{A}:\Mat{N}{\C}\to\Mat{N}{\C}$ be the bounded linear operator given by $\displaystyle \mathfrak{A}(M) = \frac{1}{N!} \sum_{\pi\in \calS_N} \pi M \pi^{-1}$. Then
\begin{equation*}
\mathfrak{A}(M) = \tr{M}{\rm I}_N + \left(\frac{1}{N(N-1)}\sum_{a\neq b} M_{a,b}\right) ({\rm J}_N - {\rm I}_N).
\end{equation*}
\end{lemma}

\begin{IEEEproof}
A straightforward computation shows that
\begin{align*}
\mathfrak{A}(M)_{i,j} &= \frac{1}{N!} \sum_{\pi\in \calS_N} \sum_{a,b=1}^N \pi_{i,a} \pi_{j,b} M_{a,b}.
\end{align*}
By definition, $\pi_{i,a} = \delta_{\pi(i)}(a)$ and $\pi_{j,b} = \delta_{\pi(j)}(b)$, thus
\begin{align*}
\mathfrak{A}(M)_{i,j} &= \frac{1}{N!} \sum_{\pi\in \calS_N} M_{\pi(i),\pi(j)}.
\end{align*}
When $i=j$, the previous sum reduces to
\begin{align*}
\mathfrak{A}(M)_{i,i} &= \frac{1}{N!} \sum_{k=1}^N M_{k,k} \sum_{\begin{smallmatrix}\pi\in \calS_N\\ \pi(i)=k\end{smallmatrix}} 1\\
&= \frac{1}{N} \sum_{k=1}^N M_{k,k}.
\end{align*}
If $i\neq j$, then
\begin{align*}
\mathfrak{A}(M)_{i,j} &= \frac{1}{N!} \sum_{a\neq b} M_{a,b}  \sum_{\begin{smallmatrix}\pi\in \calS_N\\ \pi(i)=a, \pi(j)=b\end{smallmatrix}} 1\\
&= \frac{1}{N(N-1)} \sum_{a\neq b} M_{a,b}.
\end{align*}
From this the result follows.
\end{IEEEproof}

The eigenvalues of a matrix of the form $\alpha{\rm I}_N+\beta{\rm J}_N$ 
depend on $\alpha$ and $\beta$ but its eigenvectors do not. Indeed, an
orthonormal eigenbasis for any such matrix can be obtained by applying
the Gram--Schmidt process to the eigenbasis
\begin{equation*}
\{e_1+e_2+\cdots+e_N,e_1-e_2,e_2-e_3,\ldots,e_{N-1}-e_N\},
\end{equation*}
where $\{e_1,\ldots,e_N\}$ is the standard basis in $\C^N$. Thus,
\begin{equation*}
\alpha{\rm I}_N + \beta{\rm J}_N = U_0 D(\alpha,\beta) U_0^*,
\end{equation*}
where $D(\alpha,\beta)=\diag{\alpha+N\beta,\alpha,\alpha,\cdots,\alpha}$ and $U_0$ is the unitary matrix given in Remark~\ref{Remark}, which is independent from $\alpha$ and $\beta$.

The operator $\mathfrak{A}$ appeared implicitly in the original proof of Telatar's theorem about the capacity of the point-to-point multiantenna Gaussian channel mentioned in the Introduction. Here we will use this operator twice to derive Theorem~\ref{Theorem:ClassSImpliesClassB}.

\begin{IEEEproof}[Proof of Theorem~\ref{Theorem:ClassSImpliesClassB}]
For notational simplicity, we define the mapping $\mathcal{I}:\CvM{dN}{\C}\to\R$ by
\begin{equation*}
\mathcal{I}(Q) = \E{\log\det({\rm I}_{dN}+PH_{dN} Q H_{dN}^*)}.
\end{equation*}
For a given $Q\in\CvM{dN}{\C}$ such that $\Tr{Q}=1$, we will show that there exist covariance matrices $Q'$, $Q''$ and $Q'''$ such that:
\begin{itemize}
	\item[i)] $\Tr{Q'}=\Tr{Q''}=\Tr{Q'''}=1$;
	\item[ii)] $\mathcal{I}(Q)\leq\mathcal{I}(Q')\leq\mathcal{I}(Q'')\leq\mathcal{I}(Q''')$;
	\item[iii)] $Q'''=\dfrac{1}{N}\Phi\otimes{\rm I}_N$ for some $\Phi\in\CvM{d}{\C}$.
\end{itemize}
In this way, if $Q$ attains the capacity, then there exists a covariance matrix of the intended form which also attains the capacity.

Let $Q^{(i,j)}\in\mathnormal{M}_{N}\left({\mathbb{C}}\right)$ for $1\leq i,j\leq d$ be such that $Q$ equals $(Q^{(i,j)})_{i,j=1}^d$, i.e., divide $Q$ into $d\times d$ blocks, each of size $N\times N$. Define
\begin{equation*}
Q' = \frac{1}{N!} \sum_{\pi\in \calS_N} ({\rm I}_d\otimes\pi)Q({\rm I}_d \otimes\pi)^{-1}.
\end{equation*}
Note that ${\rm I}_{dN}+PH_{dN}Q'H_{dN}^*$ equals
\begin{equation*}
{\rm I}_{dN} + PH_{dN} \left[\frac{1}{N!} \sum_{\pi\in \calS_N} ({\rm I}_d\otimes\pi)Q({\rm I}_d\otimes\pi)^{-1}\right] H_{dN}^*.
\end{equation*}
Since the function $\cdot\mapsto\log\det({\rm I}_{dN}+PH_{dN}\cdot H_{dN}^*)$ is concave,
\begin{equation*}
\frac{1}{N!} \sum_{\pi\in \calS_N} \log\det({\rm I}_{dN}+PH_{dN}({\rm I}_d\otimes\pi)Q({\rm I}_d\otimes\pi)^{-1}H_{dN}^*)
\end{equation*}
lower bounds $\log\det({\rm I}_{dN}+PH_{dN}Q'H_{dN}^*)$. Since $\pi\in\calS_N$ is a unitary matrix, by the defining condition for $\class{S}$-channels \eqref{eq:DefiningConditionSChannels}, we have that $H_{dN}\stackrel{\mathcal{L}}{=} H _{dN}({\rm I}_d\otimes \pi)$ for all $\pi\in\calS_N$. Thus
\begin{equation*}
\mathcal{I}(Q') \geq \mathcal{I}(Q).
\end{equation*}
This shows that the mutual information induced by $Q'$ is greater than or equal to the one produced by $Q$.

Observe that $(%
{\rm I}_d\otimes\pi)Q({\rm I}_d\otimes\pi)^{-1}=(\pi Q^{(i,j)} \pi^{-1})_{i,j=1}^d$, and thus $Q'=(\mathfrak{A}(Q^{(i,j)}))_{i,j=1}^d$. By the discussion following the previous lemma, we can simultaneously diagonalize all the blocks of $Q'$. In particular,
\begin{align*}
Q' = ({\rm I}_d\otimes U_0) Q'' ({\rm I}_d\otimes U_0)^*,
\end{align*}
where $Q''$ is a $d\times d$ block matrix with all its blocks being diagonal matrices of size $N\times N$. Again by the defining condition for $\class{S}$-channels,
\begin{equation*}
\mathcal{I}(Q')=\mathcal{I}(({\rm I}_d\otimes U_0)Q''({\rm I}_d\otimes U_0)^*)=\mathcal{I}(Q''),
\end{equation*}
i.e., the mutual informations induced by $Q'$ and $Q''$ are equal. Finally, define
\begin{equation*}
Q''':= \frac{1}{N!} \sum_{\pi\in \calS_N} ({\rm I}_d\otimes\pi)Q''({\rm I}_d\otimes\pi)^{-1}.
\end{equation*}
Since all the blocks of $Q''$ are diagonal matrices and $Q'''$ equals $(\mathfrak{A}(Q''^{(i,j)}))_{i,j=1}^d$, the previous lemma implies that $Q''' = \dfrac{1}{N} \Phi\otimes {\rm I}_N$ for some complex matrix  $\Phi\in\Mat{d}{\C}$. As we did with $Q'$ and $Q$, the mutual information induced by $Q'''$ is greater than or equal to that of $Q''$. It is not hard to verify that $\Phi\in\CvM{d}{\C}$ and that $\Tr{\Phi}=1$.
\end{IEEEproof}

\begin{IEEEproof}[Proof of Proposition~\ref{Proposition:AlgebraicClassS}]
Let $U_N\in\calU(N)$. Both $H_1$ and $H_2$ belong to $\class{S}_{d,N}$. Hence, $H_1 \stackrel{\mathcal{L}}{=} H_1 U_N$ and $H_2 \stackrel{\mathcal{L}}{=} H_2 U_N$. The random matrices $H_1 U_N$ and $H_2 U_N$ are independent, as $H_1$ and $H_2$ are independent and $U_N$ is constant. Therefore $(H_1+H_2)U_N = H_1 U_N + H_2 U_N \stackrel{\mathcal{L}}{=} H_1 +H_2$. In other words, $H_1+H_2\in\class{S}_{d,N}$, as required. The proof for the product follows the same lines.
\end{IEEEproof}

\begin{IEEEproof}[Proof of Theorem~\ref{Theorem:OpVCapacity}]
Define $\displaystyle L:=\limsup_{N\rightarrow \infty }\frac{C(H_{dN})}{dN}$. By definition, there exists an increasing sequence $(N_{i})_{i=1}^{\infty }\subset {\mathbb{N}}$ such that
\begin{equation*}
\lim_{i\rightarrow \infty }\frac{C_{dN_i}(H_{dN_i})}{dN_i}=L.
\end{equation*}
By assumption, $H_{dN}$ is a $\class{B}_{d,N}$-channel for all $N\in\N$. Thus, there exists a sequence $(\Phi_{N_{i}})_{i=1}^{\infty }\subset\CvM{d}{\C}$ of covariance matrices with unital trace such that $\dfrac{P}{N_i}\Phi_{N_i}\otimes {\rm I}_{N_i}$ is a capacity achieving input covariance matrix for $H_{dN_i}$, for every $i\geq1$. Since the set of $d\times d$ covariance matrices with a given trace is compact, there exists a covariance matrix $\Phi_{\infty }$ with unital trace such that
\begin{equation}
\label{eq:Qinfty}
\Phi_{\infty }=\lim_{j\rightarrow \infty }\Phi_{N_{i_{j}}}
\end{equation}
for a subsequence $(N_{i_{j}})_{j=1}^{\infty }$ of $(N_{i})_{i=1}^{\infty }$.

Let
\begin{equation*}
\left((H_{dN_{i_j}}(\Phi_{N_{i_j}}\otimes{\rm I}_{N_{i_j}})H_{dN_{i_j}}^*/N_{i_j})^{(p,q)}\right)_{p,q=1}^d
\end{equation*}
be the $d\times d$ blocks of size $N_{i_j}\times N_{i_j}$ of the $dN_{i_j}\times dN_{i_j}$ matrix
\begin{equation*}
\frac{1}{N_{i_j}}H_{dN_{i_j}}(\Phi_{N_{i_j}}\otimes{\rm I}_{N_{i_j}})H_{dN_{i_j}}^*.
\end{equation*}
Combining \eqref{eq:Qinfty} and the hypothesis that $(H_{dN}^{(p,q)})_{p,q=1}^d$ converges in $\ast$-distribution to $({\bf H}_{p,q})_{p,q=1}^d\subset\calA$, a straightforward computation shows that
\begin{equation*}
\left((H_{dN_{i_j}}(\Phi_{N_{i_j}}\otimes{\rm I}_{N_{i_j}})H_{dN_{i_j}}^*/N_{i_j})^{(p,q)}\right)_{p,q=1}^d
\end{equation*}
converges in $\ast$-distribution to $(({\bf H}\Phi_\infty{\bf H}^*)_{p,q})_{p,q=1}^d$. In particular, this implies that
\begin{equation*}
F^{H_{dN_{i_j}}(\Phi_{N_{i_{j}}}\otimes{\rm I}_{N_{i_j}})H_{dN_{i_j}}^{\ast }/N_{i_{j}}}\Rightarrow F^{\mathbf{H}\Phi_{\infty }\mathbf{H}^{\ast }},
\end{equation*}
where $F^{H_{dN_{i_j}}(\Phi_{N_{i_{j}}}\otimes{\rm I}_{N_{i_j}})H_{dN_{i_j}}^{\ast }/N_{i_{j}}}$ is the mean eigenvalue distribution of the corresponding random matrix. By a standard argument,
\begin{align*}
L &= \lim_{j\rightarrow \infty }\frac{C_{dN_{i_{j}}}(H_{dN_{i_j}})}{dN_{i_{j}}}\\
&= \lim_{j\rightarrow \infty }\int_{\mathbb{R}}\log \left( 1+ Px\right) \text{d}F^{H_{dN_{i_j}}(\Phi_{N_{i_{j}}}\otimes{\rm I}_{N_{i_j}})H_{dN_{i_j}}^{\ast }/N_{i_{j}}}(x)\\
&= \int_{\mathbb{R}}\log \left( 1+Px\right) \text{d}F^{\mathbf{H}\Phi_{\infty }\mathbf{H}^{\ast }}(x).
\end{align*}
Then, by the definition of $C({\bf H})$,
\begin{equation}
\label{eq:CHgeqL}
C({\bf H}) \geq L = \limsup_{N\rightarrow \infty }\frac{C(H_{dN})}{dN}.
\end{equation}

Let $\Phi\in\CvM{d}{\C}$. Denote by $x^{\Phi,N}$ and $y^{\Phi,N}$ the input and output vectors when the covariance of the input vector is $\frac{P}{N}\Phi\otimes{\rm I}_N$. As before, we have that the blocks of $H_{dN}(\Phi\otimes{\rm I}_N)H_{dN}^{\ast }/N$ converge in $\ast$-distribution to the entries of ${\bf H}\Phi{\bf H}^*$. Thus $F^{H_{dN}(\Phi\otimes{\rm I}_N)H_{dN}^{\ast }/N} \Rightarrow F^{\mathbf{H}\Phi\mathbf{H}^{\ast }}$ and also
\begin{equation*}
\frac{I(x^{\Phi,N};(y^{\Phi,N},H_{dN}))}{dN} \stackrel{N\to\infty}{\longrightarrow} \int_\R \log \left( 1+Px\right) \text{d}F^{\mathbf{H}\Phi\mathbf{H}^{\ast }}(x).
\end{equation*}
Let $P\Phi_0\in\Mat{d}{\C}$ be a capacity achieving input covariance matrix for the abstract capacity of ${\bf H}$. By the previous equation,
\begin{align*}
C({\bf H}) &= \int_{\mathbb{R}}\log \left( 1+Px\right) \text{d}F^{\mathbf{H}\Phi_0\mathbf{H}^{\ast }}(x)\\
&= \lim_{N\rightarrow \infty }\frac{I(x^{\Phi_0,N};(y^{\Phi_0,N},H_{dN}))}{dN}.
\end{align*}
Normalizing and taking liminf on both sides of
\begin{equation*}
C(H_{dN})\geq I(x^{\Phi_0,N};(y^{\Phi_0,N},H_{dN})),
\end{equation*}
we obtain that
\begin{align*}
\liminf_{N\rightarrow \infty }\frac{C(H_{dN})}{dN} &\geq \liminf_{N\rightarrow\infty }\frac{I(x^{\Phi_0,N};(y^{\Phi_0,N},H_{dN}))}{dN}\\
&= C(\mathbf{H)}.
\end{align*}
Combining (\ref{eq:CHgeqL}) with the previous inequality, we obtain
\begin{equation*}
\lim_{N\to\infty} \frac{C(H_{dN})}{dN} = C({\bf H}),
\end{equation*}
which completes the proof.
\end{IEEEproof}

\section{Free probability theory}
\label{Appendix:FreeProbability}

A non-commutative probability space is a pair $(\calA,\FI)$ such that $\calA$ is a unital algebra with unit ${\bf 1}_\calA$ and $\FI:\calA\to\C$ is a linear functional such that $\F{{\bf 1}_\calA}=1$. In this paper, we always assume that $\calA$ is a $C^*$-algebra and $\FI$ is tracial and positive, i.e., $\F{ab}=\F{ba}$ for all $a,b\in\calA$ and $\F{a^*a}\geq0$ for all $a\in\calA$. The elements of $\calA$ are called non-commutative random variables. In our setting, an operator-valued random variable is an element in $\Mat{n}{\calA}$ for some $n\in\N$. For a more comprehensive treatment of the theory of free probability, see Nica and Speicher \cite{NiSp06}. In the case of operator-valued free probability, see the original approach of Voiculescu \cite{Vo85} and the combinatorial approach of Speicher \cite{Sp98}.

A non-commutative random variable $s\in\calA$ is, by definition, a semicircular random variable with covariance $\sigma^2$ if and only if $s=s^*$ and for all $k\in\N$
\begin{equation*}
\F{s^k} = \begin{cases}
\sigma^k C_{k/2}  & \textnormal{if }k\textnormal{ is even,}\\
0 & \textnormal{otherwise},
\end{cases}
\end{equation*}
where $\displaystyle C_k=\frac{1}{k+1} \binom{2k}{k}$ is the $k$th Catalan number \cite{Ko09}. A non-commutative random variable $c\in\calA$ is, also by definition, a circular random variable with covariance $\sigma^2$ if and only if $\Re(c)=\dfrac{c+c^*}{2}$ and $\Im(c)=\dfrac{c-c^*}{2i}$ are free semicircular non-commutative random variables with covariance $\sigma^2/2$. See below for the definition of freeness.

Given a non-commutative random variable $a\in\calA$, its \mbox{$\ast$-distribution} is the linear functional $\mu:\C\langle X,X^*\rangle\to\C$ given by $\mu(p) = \F{p(a,a^*)}$ for any $p\in\C\langle X,X^*\rangle$ where $\C\langle X,X^*\rangle$ is the $\ast$-algebra of polynomials in the non-commutative indeterminates $X$ and $X^*$. For $a\in\calA$ selfadjoint, a compactly supported real probability distribution $F^a$ is called the spectral distribution of $a$ if for all $n\in\N$
\begin{equation*}
\F{a^n} = \int_\R t^n \textrm{d} F^a(t).
\end{equation*}
Identifying $\Mat{n}{\calA}$ with $\Mat{n}{\C}\otimes\calA$, a compactly supported real probability distribution $F^{\bf X}$ is the spectral distribution of a self-adjoint operator-valued random variable ${\bf X}={\bf X}^*$ if
\begin{equation*}
\int_\R \frac{1}{z-t} \textrm{d} F^{\bf X}(t) = (\FI\otimes\textrm{tr}) \left( (z{\rm I}_n-{\bf X})^{-1}\right).
\end{equation*}
for all $z\in\C^+=\{z\in\C : \Im(z)>0\}$.

For $\mathcal{X}\subset\calA$, let $\C\langle\mathcal{X}\rangle$ denote the unital algebra generated by $\mathcal{X}$. The families $\mathcal{X}_1,\ldots,\mathcal{X}_k\subset\calA$ are said to be free if
\begin{equation*}
\F{x_1\cdots x_n}=0
\end{equation*}
whenever $n\in\N$, $x_j\in \C\langle\mathcal{X}_{i_j}\rangle$, $i_1\neq i_2\neq \cdots \neq i_n$ and $\F{x_j}=0$ for all $1\leq j\leq n$.

For each $N\in\N$, let $A_N^{(1)},\ldots,A_N^{(k)}$ be $N\times N$ random matrices. We say that $(A_N^{(1)},\ldots,A_N^{(k)})$ converge in $\ast$-distribution to $(a_1,\ldots,a_k)\subset\calA$, denoted by
\begin{equation*}
(A_N^{(1)},\ldots,A_N^{(k)})\stackrel{\ast\textnormal{-dist}}{\longrightarrow} (a_1,\ldots,a_k),
\end{equation*}
if $\lim\limits_{N\to\infty} \E{\tr{p(A_N^{(1)},A_N^{(1)*},\cdots,A_N^{(k)},A_N^{(k)*})}}$ equals
\begin{equation*}
\F{p(a_1,a_1^*,\cdots,a_k,a_k^*)}
\end{equation*}
for all $p\in\C\langle X_1,X_1^*,\cdots,X_k,X_k^*\rangle$. Moreover, we say that $A_N^{(1)},\ldots,A_N^{(k)}$ are asymptotically free if $a_1,\ldots,a_k$ are free.

The following theorems, due to Voiculescu \cite{Vo91,Vo98}, are the earliest connections between random matrices and free probability. We restate them in the notation of the present paper for the convenience of the reader.

\begin{theorem}
\label{Thm:VoiculescuThm}
For each $N\in\N$, let $W_N^{(1)},\ldots,W_N^{(p)}$ be $p$ independent $N\times N$ Gaussian matrices and let $D_N^{(1)},\ldots,D_N^{(q)}$ be $q$ constant $N\times N$ matrices that converge in $\ast$-distribution. Then there exist a non-commutative probability space $(\calA,\FI)$ and non-commutative random variables $c_1,\ldots,c_p,d_1,\ldots,d_q\in\calA$ such that
\begin{itemize}
	\item[a)] $c_1,\ldots,c_p$ are free circular random variables,
	\item[b)] $c_i$ and $d_j$ are free for all $1\leq i\leq p$ and $1\leq j\leq q$,
	\item[c)] the family $(W_N^{(1)}/\sqrt{N},\ldots,W_N^{(p)}/\sqrt{N},D_N^{(1)},\ldots,D_N^{(q)})$ converges in $\ast$-distribution to $(c_1,\ldots,c_p,d_1,\ldots,d_q)$.
\end{itemize}
\end{theorem}

\begin{theorem}
\label{Thm:VoiculescuThm2}
Assume that $A_N^{(1)},\ldots,A_N^{(p)}$ are $N\times N$ diagonal constant real matrices with individual limits in distribution, i.e., $A_N^{(i)}$ converges in distribution for each $1\leq i\leq p$. If $U_N^{(1)},\ldots,U_N^{(p)}$ are independent $N\times N$ Haar unitary random matrices and $W_N^{(1)},\ldots,W_N^{(q)}$ are $N\times N$ Gaussian matrices independent among themselves and independent from $U_N^{(1)},\ldots,U_N^{(p)}$, then there exists a non-commutative probability space $(\calA,\FI)$ and non-commutative random variables $a_1,\ldots,a_p,c_1,\ldots,c_q\in\calA$ such that
\begin{itemize}
	\item[a)] $a_1,\ldots,a_p$ are free selfadjoint random variables,
	\item[b)] $c_1,\ldots,c_q$ are free circular random variables,
	\item[c)] $a_i$ and $c_j$ are free for all $1\leq i\leq p$ and $1\leq j\leq q$,
	\item[d)] the family
	\begin{equation*}
		(U_N^{(1)}A_N^{(1)}U_N^{(1)*},\ldots,U_N^{(p)}A_N^{(p)}U_N^{(p)*},W_N^{(1)},\ldots,W_N^{(q)})
	\end{equation*}
	converges in $\ast$-distribution to $(a_1,\ldots,a_p,c_1,\ldots,c_q)$.
\end{itemize}
\end{theorem}

Given $a\in\calA$, its Cauchy transform $G_a:\C^+\to\C^-$ is defined by $G_a(z) = \F{(z-a)^{-1}}$ for $z\in\C^+$. Denote by $\Mat{n}{\C}^+$ the set of matrices $M\in\Mat{n}{\C}$ such that
\begin{equation*}
\frac{M-M^*}{2i} >0.
\end{equation*}
The $\Mat{n}{\C}$-valued Cauchy transform $\calG_{\bf X}$ of a self-adjoint operator-valued random variable ${\bf X}$ in $\Mat{n}{\calA}$ is given, for $B\in\Mat{n}{\C}^+$, by
\begin{equation*}
\calG_{\bf X}(B) := ({\bf 1}_{\Mat{n}{\C}}\otimes\FI) \left((B-{\bf X})^{-1}\right).
\end{equation*}

If the spectral distribution of $a=a^*\in\calA$ has a density at $x\in\R$, then the Stieltjes inversion formula states that
\begin{equation*}
\frac{\textrm{d}}{\textrm{d} x} F^a(x) = -\frac{1}{\pi} \lim_{\epsilon\to0^+} \Im(G_a(x+i\epsilon)),
\end{equation*}
where $\Im$ means taking the imaginary part. Similarly, for a self-adjoint operator-valued random variable ${\bf X}$ in $\Mat{n}{\calA}$,
\begin{equation*}
\frac{\textrm{d}}{\textrm{d} x} F^{\bf X}(x) = -\frac{1}{\pi} \lim_{\epsilon\to0^+} \Im(\tr{\calG_{\bf X}((x+i\epsilon){\rm I}_n)}),
\end{equation*}
whenever $F^{\bf X}$ has a density at $x\in\R$.


\begin{thebibliography}{10}
\providecommand{\url}[1]{#1}
\csname url@samestyle\endcsname
\providecommand{\newblock}{\relax}
\providecommand{\bibinfo}[2]{#2}
\providecommand{\BIBentrySTDinterwordspacing}{\spaceskip=0pt\relax}
\providecommand{\BIBentryALTinterwordstretchfactor}{4}
\providecommand{\BIBentryALTinterwordspacing}{\spaceskip=\fontdimen2\font plus
\BIBentryALTinterwordstretchfactor\fontdimen3\font minus
  \fontdimen4\font\relax}
\providecommand{\BIBforeignlanguage}[2]{{%
\expandafter\ifx\csname l@#1\endcsname\relax
\typeout{** WARNING: IEEEtran.bst: No hyphenation pattern has been}%
\typeout{** loaded for the language `#1'. Using the pattern for}%
\typeout{** the default language instead.}%
\else
\language=\csname l@#1\endcsname
\fi
#2}}
\providecommand{\BIBdecl}{\relax}
\BIBdecl

\bibitem{FoGa98}
G.~Foschini and M.~Gans, ``On limits of wireless communications in a fading
  environment when using multiple antennas,'' \emph{Wireless Personal
  Communications}, vol.~6, no.~3, pp. 311--335, 1998.

\bibitem{Fo96}
G.~J. Foschini, ``Layered space-time architecture for wireless communication in
  a fading environment when using multi-element antennas,'' \emph{Bell Labs
  Technical Journal}, vol.~1, no.~2, pp. 41--59, 1996.

\bibitem{TuVe04}
A.~M. Tulino and S.~Verd\'{u}, ``Random matrix theory and wireless
  communications,'' \emph{Foundations and Trends® in Communications and
  Information Theory}, vol.~1, no.~1, pp. 1--182, 2004.

\bibitem{Te99}
E.~Telatar, ``Capacity of multi-antenna gaussian channels,'' \emph{European
  Transactions on Telecommunications}, vol.~10, no.~6, pp. 585--595, 1999.

\bibitem{MaPa67}
V.~A. Marchenko and L.~A. Pastur, ``Distribution of eigenvalues for some sets
  of random matrices,'' \emph{Mathematics of the USSR-Sbornik}, vol.~1, no.~4,
  pp. 457--483, 1967.

\bibitem{EdLaMaTu14}
E.~G. Larsson, O.~Edfors, F.~Tufvesson, and T.~L. Marzetta, ``Massive mimo for
  next generation wireless systems,'' \emph{IEEE Communications Magazine},
  vol.~52, no.~2, pp. 186--195, February 2014.

\bibitem{BaSi10}
Z.~Bai and J.~W. Silverstein, \emph{Spectral analysis of large dimensional
  random matrices}, 2nd~ed., ser. Springer Series in Statistics.\hskip 1em plus
  0.5em minus 0.4em\relax Springer, New York, 2010.

\bibitem{TaVu12}
T.~Tao and V.~Vu, ``Random matrices: the universality phenomenon for {W}igner
  ensembles,'' in \emph{Modern aspects of random matrix theory}, ser. Proc.
  Sympos. Appl. Math.\hskip 1em plus 0.5em minus 0.4em\relax Amer. Math. Soc.,
  Providence, RI, 2014, vol.~72, pp. 121--172.

\bibitem{TaVu12_2}
------, ``Random covariance matrices: universality of local statistics of
  eigenvalues,'' \emph{Ann. Probab.}, vol.~40, no.~3, pp. 1285--1315, 2012.

\bibitem{Di14}
X.~Ding, ``Spectral analysis of large block random matrices with rectangular
  blocks,'' \emph{Lith. Math. J.}, vol.~54, no.~2, pp. 115--126, 2014.

\bibitem{GoKoTi15}
F.~G\"otze, H.~K\"osters, and A.~Tikhomirov, ``Asymptotic spectra of
  matrix-valued functions of independent random matrices and free
  probability,'' \emph{Random Matrices Theory Appl.}, vol.~4, no.~2, pp. 1--85,
  2015.

\bibitem{LoTuVe05}
A.~M. Tulino, A.~Lozano, and S.~Verd\'u, ``Impact of antenna correlation on the
  capacity of multiantenna channels,'' \emph{IEEE Trans. Inform. Theory},
  vol.~51, no.~7, pp. 2491--2509, 2005.

\bibitem{Gi90}
V.~L. Girko, \emph{Theory of random determinants}, ser. Mathematics and its
  Applications (Soviet Series).\hskip 1em plus 0.5em minus 0.4em\relax Kluwer
  Academic Publishers Group, Dordrecht, 1990, vol.~45, translated from the
  Russian.

\bibitem{Gi01}
------, \emph{Theory of stochastic canonical equations. {V}ol. {I}}, ser.
  Mathematics and its Applications.\hskip 1em plus 0.5em minus 0.4em\relax
  Kluwer Academic Publishers, Dordrecht, 2001, vol. 535.

\bibitem{HaLoNa07}
W.~Hachem, P.~Loubaton, and J.~Najim, ``Deterministic equivalents for certain
  functionals of large random matrices,'' \emph{Ann. Appl. Probab.}, vol.~17,
  no.~3, pp. 875--930, 2007.

\bibitem{DyNiVo92}
D.~V. Voiculescu, K.~J. Dykema, and A.~Nica, \emph{Free random variables}, ser.
  CRM Monograph Series.\hskip 1em plus 0.5em minus 0.4em\relax American
  Mathematical Society, Providence, RI, 1992, vol.~1.

\bibitem{Sh96}
D.~Shlyakhtenko, ``Random {G}aussian band matrices and freeness with
  amalgamation,'' \emph{Internat. Math. Res. Notices}, no.~20, pp. 1013--1025,
  1996.

\bibitem{BrFaOrSp08}
R.~Rashidi~Far, T.~Oraby, W.~Bryc, and R.~Speicher, ``On slow-fading {MIMO}
  systems with nonseparable correlation,'' \emph{IEEE Trans. Inform. Theory},
  vol.~54, no.~2, pp. 544--553, 2008.

\bibitem{HeRaSp07}
J.~W. Helton, R.~Rashidi~Far, and R.~Speicher, ``Operator-valued semicircular
  elements: solving a quadratic matrix equation with positivity constraints,''
  \emph{Int. Math. Res. Not. IMRN}, no.~22, pp. 1--15, 2007.

\bibitem{Gi95}
V.~L. Girko, ``A matrix equation for the resolvents of random matrices with
  independent blocks,'' \emph{Teor. Veroyatnost. i Primenen.}, vol.~40, no.~4,
  pp. 741--753, 1995.

\bibitem{CoDeSi11}
R.~Couillet, M.~Debbah, and J.~W. Silverstein, ``A deterministic equivalent for
  the analysis of correlated {MIMO} multiple access channels,'' \emph{IEEE
  Trans. Inform. Theory}, vol.~57, no.~6, pp. 3493--3514, 2011.

\bibitem{SpVa12}
R.~Speicher and C.~Vargas, ``Free deterministic equivalents, rectangular random
  matrix models, and operator-valued free probability theory,'' \emph{Random
  Matrices Theory Appl.}, vol.~1, no.~2, pp. 1--26, 2012, with an appendix by
  Tobias Mai.

\bibitem{BeSpTrVa14}
S.~T. Belinschi, R.~Speicher, J.~Treilhard, and C.~Vargas, ``Operator-valued
  free multiplicative convolution: analytic subordination theory and
  applications to random matrix theory,'' \emph{Int. Math. Res. Not. IMRN},
  no.~14, pp. 5933--5958, 2015.

\bibitem{BeMaSp15}
S.~T. Belinschi, T.~Mai, and R.~Speicher, ``Analytic subordination theory of
  operator-valued free additive convolution and the solution of a general
  random matrix problem,'' \emph{Journal für die reine und angewandte
  Mathematik (Crelles Journal)}, 2015.

\bibitem{Vo85}
D.~Voiculescu, ``Symmetries of some reduced free product {$C^\ast$}-algebras,''
  in \emph{Operator algebras and their connections with topology and ergodic
  theory ({B}u\c steni, 1983)}, ser. Lecture Notes in Math.\hskip 1em plus
  0.5em minus 0.4em\relax Springer, Berlin, 1985, vol. 1132, pp. 556--588.

\bibitem{Benaych2009}
F.~Benaych-Georges, ``Rectangular random matrices, entropy, and {F}isher's
  information,'' \emph{J. Operator Theory}, vol.~62, no.~2, pp. 371--419, 2009.

\bibitem{LoTuVe06}
A.~M. Tulino, A.~Lozano, and S.~Verd\'{u}, ``Capacity-achieving input covariance
  for single-user multi-antenna channels,'' \emph{IEEE Transactions on Wireless
  Communications}, vol.~5, no.~3, pp. 662--671, March 2006.

\bibitem{Ga11}
J.~Gallier, \emph{Geometric methods and applications}, 2nd~ed., ser. Texts in
  Applied Mathematics.\hskip 1em plus 0.5em minus 0.4em\relax Springer, New
  York, 2011, vol.~38, for computer science and engineering.

\bibitem{Mu02}
R.~R. M\"uller, ``On the asymptotic eigenvalue distribution of concatenated
  vector-valued fading channels,'' \emph{IEEE Trans. Inform. Theory}, vol.~48,
  no.~7, pp. 2086--2091, 2002.

\bibitem{DeFaGeZa11}
N.~Fawaz, K.~Zarifi, M.~Debbah, and D.~Gesbert, ``Asymptotic capacity and
  optimal precoding in {MIMO} multi-hop relay networks,'' \emph{IEEE Trans.
  Inform. Theory}, vol.~57, no.~4, pp. 2050--2069, 2011.

\bibitem{Fu92}
J.~Fujii, ``Operator means and the relative operator entropy,'' in
  \emph{Operator theory and complex analysis ({S}apporo, 1991)}, ser. Oper.
  Theory Adv. Appl.\hskip 1em plus 0.5em minus 0.4em\relax Birkh\"auser, Basel,
  1992, vol.~59, pp. 161--172.

\bibitem{MaSpWe15}
T.~Mai, R.~Speicher, and M.~Weber, ``Absence of algebraic relations and of zero
  divisors under the assumption of full non-microstates free entropy
  dimension,'' \emph{Adv. Math.}, vol. 304, pp. 1080--1107, 2017.

\bibitem{Sh14}
I.~{Charlesworth} and D.~{Shlyakhtenko}, ``{Regularity of Polynomials in Free
  Variables},'' \emph{ArXiv e-prints}, Aug. 2014.

\bibitem{Sp98}
R.~Speicher, ``Combinatorial theory of the free product with amalgamation and
  operator-valued free probability theory,'' \emph{Mem. Amer. Math. Soc.}, vol.
  132, no. 627, pp. x+88, 1998.

\bibitem{Sp15}
------, ``Polynomials in asymptotically free random matrices,'' \emph{Acta
  Phys. Polon. B}, vol.~46, no.~9, pp. 1611--1624, 2015.

\bibitem{NiSp06}
A.~Nica and R.~Speicher, \emph{Lectures on the combinatorics of free
  probability}, ser. London Mathematical Society Lecture Note Series.\hskip 1em
  plus 0.5em minus 0.4em\relax Cambridge University Press, Cambridge, 2006,
  vol. 335.

\bibitem{Ko09}
T.~Koshy, \emph{Catalan numbers with applications}.\hskip 1em plus 0.5em minus
  0.4em\relax Oxford University Press, Oxford, 2009.

\bibitem{Vo91}
D.~Voiculescu, ``Limit laws for random matrices and free products,''
  \emph{Invent. Math.}, vol. 104, no.~1, pp. 201--220, 1991.

\bibitem{Vo98}
------, ``A strengthened asymptotic freeness result for random matrices with
  applications to free entropy,'' \emph{Internat. Math. Res. Notices}, no.~1,
  pp. 41--63, 1998.

\end{thebibliography}


\vfill

\begin{IEEEbiographynophoto}{Mario Diaz} is currently a Ph.D. candidate in the Mathematics and Statistics Department at Queen's University, Canada. He received his B.Eng. degree in Electrical Engineering in 2011 from the University of Guadalajara, Mexico, and his M.Sc. degree in Probability and Statistics in 2013 from the Center for Research in Mathematics (CIMAT), Mexico. His research interests include random matrix theory, free probability theory, analysis of the capacity of multiantenna channels, and information privacy from a mathematical and statistical perspective. Mario Diaz is a recipient of the Ontario Trillium Scholarship.
\end{IEEEbiographynophoto}

\begin{IEEEbiographynophoto}{Victor P\'{e}rez-Abreu} received the B.Sc. degree in Physics and Mathematics from the National Polytechnic Institute, Mexico, in 1997 and the Ph.D. in Statistics in 1985 from the University of North Carolina at Chapel Hill, USA. Since 1986 he is a Research Fellow at Center for Research in Mathematics (CIMAT), in Guanajuato, Mexico. He has published some 60 papers in probability, statistics and stochastic processes. He served as founder chair of the statistics department at CIMAT, general director of CIMAT, President of the Bernoulli Society for Probability and Statistics, and Vice President of the International Statistical Institute. He was member of the steering committee of the International Year of Statistics in 2013. He is a fellow of the American Statistical Association and the Institute of Mathematical Statistics.
\end{IEEEbiographynophoto}

\vfill

\end{document}